\documentclass[amsmath, superscriptaddress, amssymb, aps, prd, longbibliography,nofootinbib,onecolumn, notitlepage,10pt]{revtex4-2}
\usepackage{amsmath,amssymb,mathtools}
\usepackage{natbib}
\usepackage[colorlinks=true,linkcolor=blue,citecolor=blue]{hyperref}
\usepackage{cleveref}
\usepackage[normalem]{ulem}
\usepackage{subcaption}

\usepackage{braket}
\usepackage{xcolor}
\usepackage{graphicx}
\usepackage{bm}
\labelformat{appendix}{Appendix #1}
\usepackage{comment}

\let\calccommentout\iffalse 
\let\calcshow\iftrue 

\crefname{section}{Section}{Sections}
\crefname{subsection}{Section}{Sections}
\crefname{subsubsection}{Section}{Sections}

\crefname{appendix}{Appendix}{Appendices}
\labelformat{equation}{Eq.~(#1)} 
\labelformat{figure}{Fig.~#1} 
\labelformat{subfigure}{Fig.~\thefigure#1} 
\labelformat{table}{Tab.~#1} 
\labelformat{appendix}{Appendix #1}

\begin{document}
\title{Correlation Harvesting by Unruh--DeWitt Detectors in a Cylindrical
Cavity: Entanglement, Mutual Information, and Quantum Discord}
\author{Shagun Kaushal}
\email{shagun.kaushal@vit.ac.in}
\affiliation{Department of Physics, School of Advanced Sciences, Vellore Institute of Technology, Vellore 632014, India}

\author{Harkirat Singh Sahota}
\email{harkirat221@gmail.com}
\affiliation{Centre for Strings, Gravitation and Cosmology, Department of Physics, Indian Institute of Technology Madras, Chennai 600036, India}

\date{\today}
\begin{abstract}

We study correlations between two Unruh--DeWitt detectors coupled to a massless scalar field confined inside a cylindrical cavity through entanglement negativity, mutual information and quantum discord measures. The entanglement negativity follows the free space behavior for small energy gaps and small separation, while it gets enhanced for large energy gap as the off-axis detector reach cavity boundary. The entanglement island, where negativity is nonzero, gets extended for large energy gaps as compared to free space and the detectors can get entangled for large separations when one of them is placed near cavity boundary. In contrast, quantum mutual information and quantum discord remain nonzero over a substantially broader parameter range, including regions in which the negativity has vanished. However, these correlations are suppressed inside the cavity as compared to the free space correlations. Therefore, boundary conditions can selectively reshape the structure of harvested correlations in quantum field theory.

\end{abstract}

\maketitle
\section{Introduction}
Quantum entanglement is one of the defining features of quantum theory \cite{bell_1, bell_2, bell_3, bell_4, WJHN, DAC}. In quantum field theory, the vacuum state itself exhibits nonzero quantum correlations owing to the Heisenberg uncertainty principle \cite{Reynaud:2001kc, Sakharov:1967pk, Streeruwitz:1975wzf}. Entanglement harvesting refers to the process by which localized quantum systems, interacting locally with a quantum field, can extract these ubiquitous correlations and become entangled without any direct interaction \cite{Menezes:2015veo, Menezes:2015iva, Lindel:2023rfi, Lima:2023pyt}. While most studies focus on harvested entanglement, detector-field interaction can also generate broader classes of correlations between the quantum systems. In particular, it is natural to ask how more general measures of correlations, quantum discord and mutual information, which characterize non-classical and total correlations, respectively, emerge in the harvesting protocol. Mutual information and discord harvesting have previously been investigated in several relativistic and curved spacetime scenarios, including accelerating detectors and black hole backgrounds~\cite{Gallock-Yoshimura:2021xsy, Bueley:2022ple, Quan:2026rmk, Lin:2024roh}.

Detector-based extraction of vacuum correlations has been extensively studied using Unruh-DeWitt detectors in both flat and curved spacetime settings, where spacetime geometry, detector motion, and causal structure significantly influence harvested correlations \cite{Perche:2022ykt, Perche:2023nde, deRamon:2023qcp, Elghaayda:2023igv, Barman:2022xht, Kaushal:2024zfi, deSLTorres:2023ujd, Martin-Martinez:2015qwa, Pozas-Kerstjens:2015gta, Bozanic:2023okm, Bhattacharya:2022ahn, Suryaatmadja:2022quq}.  In flat spacetime, correlations are distillable to entanglement only for finite separation and smaller energy gap as compared to the detector switching time. Since the reduced detector state is intrinsically mixed, it is natural to ask whether more general nonclassical correlations, such as quantum discord, can persist even in regimes where entanglement vanishes~\cite{Ollivier:2001fdq, Werlang:2009tvi}. Recent investigations have further explored experimentally motivated realizations in superconducting circuits, demonstrating that harvested entanglement can predominantly originate from vacuum field correlations even for connected or disconnected detectors \cite{Teixido-Bonfill:2025wqb}. Analogue gravity realizations using ultracold atoms in optical lattices have also been proposed to probe Unruh-like quantum correlations and entanglement harvesting phenomena in laboratory settings \cite{Lopez-Raven:2025ehf}. In realistic experiments, however, quantum fields are often confined within cavities \cite{Haroche2006, Walther2006, Raimond2001}, where boundary conditions discretize field modes and substantially modify detector response and vacuum correlations \cite{Lochan:2019osm, Soda:2021sql, Sahota:2026imu, Stargen:2021vtg}.

These observations raise a natural question: how do boundary
conditions reshape the hierarchy between entanglement and more
general quantum correlations? While cavity and boundary effects
on harvested entanglement have been investigated in diverse
settings---including reflecting boundaries, accelerated detectors,
multipartite coherence harvesting, and directional quantum
steering~\cite{Liu:2020jaj, Li:2024dvs, Barman:2023wkr, Ma:2024tyn,
Wu:2026woq, Salomaa:2026xzn, Mendez-Avalos:2022obb,
Huang:2025gme, Mukherjee:2023bnt}---most studies focus
primarily on entanglement measures such as negativity.
In particular, a recent analytical and numerical investigation
of entanglement harvesting in cylindrical cavities demonstrated
a strong dependence on cavity geometry, mode parity, and
distinct parameter scalings inside and outside the light
cone~\cite{Strohle:2026gdr}.
For finite-size detectors in a cavity of finite length, the
entanglement negativity was found to depend only weakly on
the cavity radius while exhibiting a stronger dependence on
the cavity length.
These findings motivate a closer examination of how cylindrical
confinement modifies harvested correlations relative to free
space, particularly whether boundary-induced changes in
entanglement are accompanied by similar changes in quantum
mutual information and discord.

In this study, we investigate correlation harvesting by two
Unruh--DeWitt detectors coupled to a massless scalar field
confined inside a cylindrical cavity. We characterize the
harvested correlations using entanglement negativity, quantum
mutual information, and quantum discord.
Our analysis reveals a clear hierarchy: depending on the
detector parameters, the leading-order negativity either
vanishes beyond a finite detector separation or decreases
continuously toward zero as the off-axis detector approaches
the cavity boundary.
For the parameters considered, confinement can extend the
region of non-zero negativity beyond the corresponding
free-space threshold.
In contrast, mutual information and quantum discord persist
over a broader parameter range, including regions where
the negativity vanishes.
However, both quantities remain
smaller inside the cavity than in free space.
The cavity-modified mode structure suppresses the nonlocal
contribution responsible for entanglement harvesting with
increasing detector separation, while the single-excitation
cross-correlation remains comparatively robust away from
the cavity wall. As the off-axis detector approaches the
Dirichlet boundary, both mutual information and quantum
discord are increasingly suppressed.
These results demonstrate that confinement affects different
classes of harvested correlations in quantitatively distinct
ways and provides a controlled setting for investigating
quantum correlations beyond entanglement.

The paper is organized as follows. In Section \ref{Theoretical Framework}, we introduce the theoretical framework, including the Unruh-DeWitt detector model and the quantization of the scalar field inside a cylindrical cavity. In Section \ref{Correlations between the detectors}, we present the formalism for computing the relevant correlation measures, namely negativity, mutual information, and quantum discord, and derive the expressions for the local and non-local contributions. In Section \ref{Section4}, we present our numerical results, analyzing the behavior of correlations as functions of the dimensionless parameters $\sigma/R$, $\rho_0/\sigma$, and $\sigma\Omega$, and discuss the emergence of sudden death of entanglement alongside the persistence of discord and total correlations. Finally, in Section \ref{Section5}, we summarize our findings and discuss their physical implications and possible extensions. 
Additional technical details, including intermediate
derivations and computational methods, are presented
in the appendices.
\section{Theoretical Framework}
\label{Theoretical Framework}
\subsection{Detector model and configuration}

In this work, we consider two identical
Unruh--DeWitt detectors, labeled A and B,
interacting locally with a massless scalar
field confined inside an
infinitely long cylindrical
cavity of radius $R$ with a perfectly
reflecting boundary. The detectors are modeled as two-level quantum systems with ground state $|g\rangle$, excited state $|e\rangle$, and energy gap $\Omega$. 

The interaction Hamiltonian for the composite detector--field system is
\begin{equation}
\label{hamiltonian1}
\hat{H}_{I}
=
\sum_{j=A,B}
\lambda_j(\tau_j)\,
\hat{\mu}_j(\tau_j)\,
\hat{\phi}\!\left[x_j(\tau_j)\right],
\end{equation}
where $j$ labels the two detectors, $A$ and $B$, and
$\lambda_j(\tau_j)$ denotes the time-dependent switching function,
including the detector--field coupling strength. The interaction-picture
monopole operator of detector $j$ is
\begin{equation}
\label{momentR}
\hat{\mu}_j(\tau_j)
=
|e_j\rangle\langle g_j|\,e^{i\Omega_j\tau_j}
+
|g_j\rangle\langle e_j|\,e^{-i\Omega_j\tau_j},
\end{equation}
where $|g_j\rangle$ and $|e_j\rangle$ denote the ground and excited
states of detector $j$, respectively, and $\Omega_j$ is its energy gap.
The detectors are initially prepared in their ground states, while the
field is in the cavity vacuum:
\begin{equation}
|\Psi_0\rangle
=
|g_Ag_B\rangle\otimes|0_{\mathrm{cav}}\rangle.
\label{initial-state}
\end{equation}

The detectors are assumed to be pointlike and at rest inside the cavity,
following stationary worldlines in cylindrical coordinates,
\begin{equation}
x_A(\tau_A)
=
\bigl(t_A(\tau_A),\rho_A,\theta_A,z_A\bigr),
\qquad
x_B(\tau_B)
=
\bigl(t_B(\tau_B),\rho_B,\theta_B,z_B\bigr),
\end{equation}
where $\tau_A$ and $\tau_B$ are their respective proper times. For
stationary detectors, $t_j(\tau_j)=\tau_j$. We place the detectors along
the same radial line and in the same longitudinal plane, so that
$\theta_A=\theta_B$ and $z_A=z_B$. Detector $A$ is located on the
cavity axis, $\rho_A=0$, while detector $B$ is positioned at a radial
distance $\rho_0$, such that $\rho_B=\rho_0$.

Having specified the detector configuration, the correlations harvested by the detectors are determined by the field two-point function. In the presence of boundaries, this correlation function is modified by the cavity geometry. We therefore compute the Wightman function of the scalar field inside the cylindrical cavity.
\subsection{Wightman function of the scalar field in a cylindrical cavity}
\label{subsec:wightman}

We consider a massless scalar field confined inside a perfectly reflecting cylindrical cavity of radius $R$, satisfying Dirichlet boundary conditions at the cavity wall,
\begin{equation}
\phi(\rho=R,\theta,z,t)=0 .
\end{equation}

The presence of the boundary discretizes the radial spectrum of the field through the zeros of the Bessel functions, while the longitudinal momentum remains continuous. As a result, the vacuum correlations differ from those in free space due to the cavity geometry. Expanding the field in cylindrical cavity modes, the positive-frequency Wightman function can be written as \cite{Stargen:2021vtg}
\begin{align}
\label{Wightmanij}
\mathcal{W}(x_i,x'_j)
=
\frac{1}{(2\pi R)^2}
\sum_{m,n}
\frac{J_m(\xi_{mn}\rho_i/R)J_m(\xi_{mn}\rho'_j/R)}
{J_{|m|+1}^2(\xi_{mn})}
\int_{-\infty}^{\infty}
\frac{\mathrm{d}k_z}{\omega_{mn}}
e^{-i\omega_{mn}(\tau_i-\tau'_j)}
e^{im(\theta_i-\theta'_j)}
e^{ik_z(z_i-z'_j)} ,
\end{align}

where $\xi_{mn}$ denotes the $n$-th zero of the Bessel function $J_m$, and 
\[
\omega_{mn}=\sqrt{k_z^2+\frac{\xi_{mn}^2}{R^2}} .
\]

For the detector configuration considered here,  the angular and longitudinal phase factors cancel, and the Wightman function then simplifies to
\begin{align}
\label{Wightman}
    \mathcal{W}(x_i,x'_j)=\frac{1}{(2\pi R)^2}\sum_{m,n}\frac{J_m(\xi_{mn}\rho_i/R)J_m(\xi_{mn}\rho'_j/R)}{J_{|m|+1}^2(\xi_{mn})}\int_{-\infty}^\infty\frac{d k_z}{\omega_{mn}}e^{-i\omega_{mn}\tau_-^{ij}}
\end{align}
where $\tau_-^{ij}=\tau_i-\tau'_j$.
Performing the longitudinal momentum integral
(see ~\ref{app:kz_integral}), the Wightman function can be
expressed as
\begin{align}
\label{Wightman_master}
\mathcal{W}(x_i,x'_j)
=
-\frac{i}{4\pi R^2}
\sum_{m,n}
\frac{
J_m(\xi_{mn}\rho_i/R)
J_m(\xi_{mn}\rho'_j/R)
}{
J_{|m|+1}^2(\xi_{mn})
}
\Bigg[
\theta(\tau_-^{ij})
H_0^{(2)}
\!\left(
\frac{\xi_{mn}}{R}\tau_-^{ij}
\right)
-
\theta(-\tau_-^{ij})
H_0^{(1)}
\!\left(
-\frac{\xi_{mn}}{R}\tau_-^{ij}
\right)
\Bigg].
\end{align}

Here, $H_0^{(1)}$ and $H_0^{(2)}$ denote the Hankel functions of
the first and second kinds, respectively, defined by
\begin{equation*}
H_0^{(1)}(z)=J_0(z)+iY_0(z),
\qquad
H_0^{(2)}(z)=J_0(z)-iY_0(z),
\end{equation*}
where $J_0$ and $Y_0$ are the Bessel functions of the first and
second kinds, respectively. This expression captures the modification of the vacuum correlations
induced by the cavity geometry, with the discrete radial modes encoded
in the Bessel zeros $\xi_{mn}$.

As detector $A$ is located at the symmetry axis, only the $m=0$ modes contribute because $J_m(0)=\delta_{m0}$. The pullback of Wightman function on detectors $A$ and $B$ trajectories (nonlocal case) reduces to
\begin{align}
 \label{WAB}
     \mathcal{W}(x_A,x_B')=-\frac{i}{4\pi R^2}\sum_{n}\frac{J_0(\xi_{0n}\eta)}{J_{1}^2(\xi_{0n})}\left[
\theta(\tau^{AB}_-)H_0^{(2)}\Big(\frac{\xi_{0n}}{R}\tau^{AB}_-\Big)
-\theta(-\tau^{AB}_-)H_0^{(1)}\Big(-\frac{\xi_{0n}}{R}\tau^{AB}_-\Big)
\right]
\end{align}
The pullbacks of the Wightman function along the individual detector worldlines take the form \ref{WAB}:
\begin{align}
    \label{WAA}
    \mathcal{W}(x_A,x_A')=
    -\frac{i}{4\pi R^2}\sum_n\frac{1}{J_1^2(\xi_{0n})}\left[
\theta(\tau^{AA}_-)H_0^{(2)}\Big(\frac{\xi_{0n}}{R}\tau^{AA}_-\Big)
-\theta(-\tau^{AA}_-)H_0^{(1)}\Big(-\frac{\xi_{0n}}{R}\tau^{AA}_-\Big)
\right]
\end{align}

\begin{align}
\label{WB}
\mathcal{W}(x_B,x_B')
=-\frac{i}{4\pi R^2}\sum_{m,n}
\frac{\left(J_m(\xi_{mn}\eta)\right)^2}{J_{|m|+1}^2(\xi_{mn})}
\Big[
\theta(\tau_-^{BB})H_0^{(2)}\!\left(\frac{\xi_{mn}}{R}\tau_-^{BB}\right)
-\theta(-\tau_-^{BB})H_0^{(1)}\!\left(-\frac{\xi_{mn}}{R}\tau_-^{BB}\right)
\Big]
\end{align}

\section{Correlation Measures and Detector Response}
\label{Correlations between the detectors}
Having obtained the Wightman function, we now compute the detector response and the correlations harvested from the field. In the Unruh-DeWitt framework, correlations between detectors arise from their local interaction with vacuum fluctuations, a process known as entanglement harvesting.

We assume the detectors are initially in their ground states and the field is in the vacuum,
\begin{equation}
        \label{instateR}
        |\text{in}\rangle=|0 \, g_A  \, g_B\rangle
    \end{equation}
which is an unentangled product state of two detectors and the field.

The time evolution in the interaction picture is given by
    \begin{equation}
        \label{outdefR}     |\text{out}\rangle=U|\text{in}\rangle=T e^{-i\int d\tau_j H_{I}(t(\tau_j))}|0\, g_A\, g_B\rangle
        \end{equation}
where $T$ denotes time ordering. Expanding perturbatively in the coupling strength,
\begin{equation}
U = I + U^{(1)} + U^{(2)} + \mathcal{O}(\lambda^3),
\end{equation}
with
\begin{eqnarray}
    \label{Uexpn}
    U^{(1)}&=&-i \int_{-\infty}^{\infty}d\tau_j\; H_{I}(t(\tau_j))\\
     U^{(2)}&=&- \int_{-\infty}^{\infty}d\tau_i\int_{-\infty}^{\tau_i}d\tau'_j\; H_{I}(t(\tau_i))  H_{I}(t^\prime(\tau_j'))
\end{eqnarray}
The corresponding density operator is
    \begin{equation}
        \label{densityoutR}
        \rho=|\text{out}\rangle \langle\text{out}|=\rho^{(0)}+\rho^{(1)}+\rho^{(2)}+O(\lambda^3)
    \end{equation}
where $\rho^{(n)}$ is of the order of $\lambda^{n}$. Tracing over the field degrees of freedom yields the reduced density matrix of the two detectors, 
    \begin{equation}
        \label{rhoABR}
       \rho_{\text{\small AB}}
={\rm Tr}_{\phi} \rho= \left(\begin{array}{cccc}
            1-X_{AA}-X_{BB} &0&0&M^*_{AB}  \\
            0&X_{AA}&X_{AB}&0 \\
        0&X^*_{AB}&X_{BB}&0 \\
        M_{AB}&0&0&0
        \end{array}\right )
    \end{equation}
written in the basis

\begin{equation}
\left\{
|g_Ag_B\rangle,\,
|e_Ag_B\rangle,\,
|g_Ae_B\rangle,\,
|e_Ae_B\rangle
\right\}.
\label{detector-basis}
\end{equation} 

The coefficients $X_{ij}$ and $M_{ij}$ are determined by the
Wightman function as follows.
For the stationary detector configuration considered here,
the detector proper times coincide with the common coordinate time,
and the two contributions to the double-excitation amplitude obtained
by exchanging the detector labels are equal.
\begin{align}
\label{Xij}
X_{ij} &=
\int_{-\infty}^{\infty} d\tau_i
\int_{-\infty}^{\infty} d\tau_j' \,
e^{-i\Omega(\tau_i-\tau_j')}
\lambda(\tau_i)\lambda(\tau_j')
\mathcal{W}_{ij}(x_i,x_j'), \\
\label{Mij}
M_{ij} &=
-2
\int_{-\infty}^{\infty} d\tau_i
\int_{-\infty}^{\tau_i} d\tau_j' \,
e^{i\Omega(\tau_i+\tau_j')}
\lambda(\tau_i)\lambda(\tau_j')
\mathcal{W}_{ij}(x_i,x_j').
\end{align}
Here $M_{ij}$ denotes the full double-excitation
coherence for distinct detectors, including both detector orderings.

Introducing the variables
$\tau_-^{ij}=\tau_i-\tau_j'$ and
$\tau_+^{ij}=(\tau_i+\tau_j')/2$, the time-ordered integration
domain $\tau_j'\leq\tau_i$ corresponds to $\tau_-^{ij}\geq0$.
This allows us to rewrite the integrals as
\begin{align}
\label{Xij+-}
X_{ij} &=
\int_{-\infty}^{\infty} d\tau_+^{ij}
\int_{-\infty}^{\infty} d\tau_-^{ij}\,
\lambda\!\left(\tau_+^{ij}+\frac{\tau_-^{ij}}{2}\right)
\lambda\!\left(\tau_+^{ij}-\frac{\tau_-^{ij}}{2}\right)
e^{-i\Omega\tau_-^{ij}}
\mathcal{W}_{ij}(x_i,x_j'), \\
\label{Mij+-}
M_{ij} &=
-2
\int_{-\infty}^{\infty} d\tau_+^{ij}
\int_{0}^{\infty} d\tau_-^{ij}\,
\lambda\!\left(\tau_+^{ij}+\frac{\tau_-^{ij}}{2}\right)
\lambda\!\left(\tau_+^{ij}-\frac{\tau_-^{ij}}{2}\right)
e^{2i\Omega\tau_+^{ij}}
\mathcal{W}_{ij}(x_i,x_j').
\end{align}
The finite time operation of the detectors is modeled with Gaussian switching function
\begin{equation}
\lambda(\tau)=
\lambda\exp\!\left(-\frac{\tau^2}{2\sigma^2}\right),
\end{equation}
where $\sigma$ sets the interaction duration.
The product of the switching functions becomes
\begin{equation}
\lambda\!\left(\tau_+^{ij}+\frac{\tau_-^{ij}}{2}\right)
\lambda\!\left(\tau_+^{ij}-\frac{\tau_-^{ij}}{2}\right)
=
\lambda^2
\exp\!\left[
-\frac{(\tau_+^{ij})^2}{\sigma^2}
-\frac{(\tau_-^{ij})^2}{4\sigma^2}
\right].
\end{equation}
For stationary trajectories, the Wightman function depends
only on $\tau_-^{ij}$.
Performing the Gaussian integral over $\tau_+^{ij}$ yields
\begin{align}
\label{Xij+-'}
X_{ij} &=
\sqrt{\pi}\sigma\,\lambda^2
\int_{-\infty}^{\infty} d\tau_-^{ij}\,
e^{-\frac{(\tau_-^{ij})^2}{4\sigma^2}}
e^{-i\Omega\tau_-^{ij}}
\mathcal{W}_{ij}(x_i,x_j'), \\
\label{Mij+-'}
M_{ij} &=
-2\sqrt{\pi}\sigma\,\lambda^2
e^{-\Omega^2\sigma^2}
\int_{0}^{\infty} d\tau_-^{ij}\,
e^{-\frac{(\tau_-^{ij})^2}{4\sigma^2}}
\mathcal{W}_{ij}(x_i,x_j').
\end{align}
The quantities $X_{AA}$ and $X_{BB}$ represent the local excitation probabilities of detectors $A$ and $B$, respectively. The term $X_{AB}$ encodes field-induced cross-correlations, corresponding to coherence between the single-excitation states of the two detectors. In contrast, $M_{AB}$ captures the nonlocal correlations responsible for entanglement harvesting. Together, these coefficients fully determine the reduced density matrix \eqref{rhoABR} and control the interplay between local noise and nonlocal correlations. In the following subsection, we use this reduced density matrix to quantify the entanglement generated between the detectors.

\subsection{Entanglement Negativity}
\label{subsec: Entanglement Negativity}
To quantify entanglement harvesting, we compute the negativity of the reduced density matrix \ref{rhoABR}. Negativity is defined as the absolute sum of the negative eigenvalues of the partially transposed density matrix $\rho_{AB}^{\text{T}_A}$,  where the partial transpose is taken with respect to subsystem $A$ \cite{NielsenChuang, Vidal:2002zz, Plenio:2005}. 

The partial transpose of $\rho_{AB}$ reads
\begin{equation}
(\rho_{AB})^{T_A}
=
\begin{pmatrix}
1-X_{AA}-X_{BB} & 0 & 0 & X_{AB}\\
0 & X_{AA} & M_{AB}^{*} & 0\\
0 & M_{AB} & X_{BB} & 0\\
X_{AB}^{*} & 0 & 0 & 0
\end{pmatrix}.
\label{rhoABTR'}
\end{equation}
The eigenvalues of $(\rho_{AB})^{T_A}$ are
\begin{align}
e_{\pm} &= \frac{1 - X_{AA} - X_{BB}}{2}
\pm \frac{1}{2}\sqrt{(1 - X_{AA} - X_{BB})^2 + 4|X_{AB}|^2}, \\
e'_{\pm} &= \frac{X_{AA} + X_{BB}}{2}
\pm \frac{1}{2}\sqrt{(X_{AA} - X_{BB})^2 + 4|M_{AB}|^2}.
\end{align}
A negative eigenvalue signals entanglement according to the Peres–Horodecki criterion \cite{Horodecki:1997vt}. At order $\mathcal{O}(\lambda^2)$, the only eigenvalue that
can contribute to the negativity is $e'_-$. The negative
contribution from $e_-$ in the truncated matrix is of order
$\mathcal{O}(\lambda^4)$ and cannot be interpreted without
the omitted fourth-order terms. Entanglement is therefore present when
\begin{equation}
\label{cond}
\begin{split}
(X_{AA}+X_{BB})^2<(X_{AA}-X_{BB})^2+4|M_{AB}|^2\;\;\Rightarrow\;\;
|M_{AB}|^2 > X_{AA}X_{BB} 
\end{split}
\end{equation}
The corresponding negativity is defined as
\begin{equation}
\label{Neg}
\mathcal{N}=\max[0,-e'_-].
\end{equation}
In the weak-coupling regime, retaining terms up to
$\mathcal{O}(\lambda^2)$ yields
\begin{equation}
\label{NegApprox}
\mathcal{N}
=
\max\!\left[
0,\,
\frac{\sqrt{(X_{AA}-X_{BB})^2+4|M_{AB}|^2}
-X_{AA}-X_{BB}}{2}
\right]
+\mathcal{O}(\lambda^4).
\end{equation}
All negativity results below are evaluated through
$\mathcal{O}(\lambda^2)$; fourth-order corrections are not included. For two-qubit states, nonzero negativity is a necessary and
sufficient criterion for entanglement \cite{Vidal:2002zz, Plenio:2005}. This expression retains the generally unequal local noise
contributions of the on-axis and off-axis detectors.
Entanglement is present when
$|M_{AB}|^2>X_{AA}X_{BB}$.
This condition shows that entanglement emerges only when the
nonlocal correlations encoded in $M_{AB}$ overcome the local
noise contributions $X_{AA}$ and $X_{BB}$.

\subsection{Quantum Discord and Mutual Information}
\label{subsec:discord}

Entanglement does not exhaust the correlations contained in a mixed
bipartite quantum state. We therefore also consider the quantum mutual
information, which quantifies the total correlations between the detectors,
and quantum discord, which characterizes the part of these correlations
that cannot be extracted by a local measurement on one subsystem
\cite{Ollivier:2001fdq}.

When the local measurement is performed on detector $A$, the quantum
discord is defined as
\begin{equation}
\mathcal{D}(B|A)
=
I(A:B)-J(B|A),
\label{discord}
\end{equation}
where
\begin{equation}
I(A:B)
=
S(\rho_A)+S(\rho_B)-S(\rho_{AB})
\label{IAB}
\end{equation}
is the quantum mutual information, and
\begin{equation}
J(B|A)
=
S(\rho_B)
-
\min_{\{\Pi_s^A\}}
S(B|\{\Pi_s^A\})
\label{CC}
\end{equation}
is the classical correlation obtained by performing a local projective
measurement $\{\Pi_s^A\}$ on detector $A$. 
Here,
\begin{equation}
S(\rho)
=
-\operatorname{Tr}(\rho\log\rho)
\end{equation}
is the von Neumann entropy, and $S(B|\{\Pi_s^A\})$ is the corresponding
conditional entropy of detector $B$.

Tracing the appropriate detector from~\ref{rhoABR} gives
\begin{equation}
\rho_A
=
\begin{pmatrix}
1-X_{AA} & 0\\
0 & X_{AA}
\end{pmatrix},
\qquad
\rho_B
=
\begin{pmatrix}
1-X_{BB} & 0\\
0 & X_{BB}
\end{pmatrix}.
\label{reduced-states}
\end{equation}
Their entropies are therefore
\begin{equation}
S(\rho_A)=H(X_{AA}),
\qquad
S(\rho_B)=H(X_{BB}),
\label{SA}
\end{equation}
where
\begin{equation}
H(x)
=
-x\log x-(1-x)\log(1-x)
\end{equation}
is the binary entropy.

The reduced detector state has been calculated
only to the order of $O(\lambda^2)$. In particular,
the double-excitation population
$\rho_{ee,ee}$ is of order $O(\lambda^4)$
and is not determined within the present
calculation.
Although $M_{AB}$ contributes directly
to the leading-order negativity, its omission
does not affect the leading-order mutual
information or quantum discord.
We therefore evaluate these entropic quantities
using a positive approximate state constructed
from the second-order response coefficients
by omitting $M_{AB}$.
The resulting expressions reproduce the
leading-order entropic correlations, while
higher-order corrections require knowledge
of the reduced detector state beyond
$O(\lambda^2)$.
The state used for the entropic quantities
is consequently
\begin{equation}
\widetilde{\rho}_{AB}^{(2)}
=
\begin{pmatrix}
1-X_{AA}-X_{BB} & 0 & 0 & 0\\
0 & X_{AA} & X_{AB} & 0\\
0 & X_{AB}^{*} & X_{BB} & 0\\
0 & 0 & 0 & 0
\end{pmatrix}.
\label{rho-entropic}
\end{equation}

The nonvanishing eigenvalues of
$\widetilde{\rho}_{AB}^{(2)}$ are
\begin{align}
\alpha_1
&=
1-X_{AA}-X_{BB},
\nonumber\\
\alpha_{3,4}
&=
\frac{1}{2}
\left[
X_{AA}+X_{BB}
\pm
\sqrt{
(X_{AA}-X_{BB})^2+4|X_{AB}|^2
}
\right].
\label{rhoAB-eigenvalues}
\end{align}

The remaining eigenvalue vanishes by
construction in $\widetilde{\rho}_{AB}^{(2)}$.
The corresponding double-excitation population
in the physical reduced state first appears
at $O(\lambda^4)$.
The joint entropy of the approximate state is
therefore
\begin{equation}
S(\widetilde{\rho}_{AB}^{(2)})
=
-\sum_{i=1,3,4}\alpha_i\log\alpha_i,
\label{joint-entropy}
\end{equation}
and the mutual information becomes
\begin{equation}
I(A:B)
=
H(X_{AA})+H(X_{BB})
+
\sum_{i=1,3,4}\alpha_i\log\alpha_i.
\label{IAB1}
\end{equation}
We next calculate the measurement-dependent conditional entropy. A
general orthonormal projective-measurement basis for detector $A$ may be
written as
\begin{align}
|\Pi_+^A\rangle
&=
\cos\frac{\theta}{2}|g_A\rangle
+
e^{i\phi}\sin\frac{\theta}{2}|e_A\rangle,
\nonumber\\
|\Pi_-^A\rangle
&=
\sin\frac{\theta}{2}|g_A\rangle
-
e^{i\phi}\cos\frac{\theta}{2}|e_A\rangle.
\label{measurement-basis}
\end{align}
The corresponding projectors are
\begin{equation}
\Pi_s^A
=
|\Pi_s^A\rangle\langle\Pi_s^A|,
\qquad s=\pm.
\end{equation}

The normalized conditional state of detector B is
\begin{equation}
\rho_{B|s}
=
\frac{
\operatorname{Tr}_A
\left[
(\Pi_s^A\otimes I_B)
\widetilde{\rho}_{AB}^{(2)}
(\Pi_s^A\otimes I_B)
\right]
}{
p_s
},
\label{conditional-state-definition}
\end{equation}

where
\begin{equation}
p_s
=
\operatorname{Tr}
\left[
(\Pi_s^A\otimes I_B)
\widetilde{\rho}_{AB}^{(2)}
\right]
\label{measurement-probability}
\end{equation}
is the probability of obtaining outcome $s$.

Using~\ref{rho-entropic}, the conditional states take the form
\begin{equation}
\rho_{B|s}
=
\frac{1}{p_s}
\begin{pmatrix}
A_s & C_s\\
C_s^* & B_s
\end{pmatrix},
\qquad
p_s=A_s+B_s,
\label{conditional-state}
\end{equation}
where
\begin{align}
A_+
&=
\cos^2\frac{\theta}{2}
(1-X_{AA}-X_{BB})
+
\sin^2\frac{\theta}{2}X_{AA},
\nonumber\\
A_-
&=
\sin^2\frac{\theta}{2}
(1-X_{AA}-X_{BB})
+
\cos^2\frac{\theta}{2}X_{AA},
\label{Apm}
\end{align}
\begin{equation}
B_+
=
\cos^2\frac{\theta}{2}X_{BB},
\qquad
B_-
=
\sin^2\frac{\theta}{2}X_{BB},
\label{Bpm}
\end{equation}
and
\begin{equation}
C_\pm
=
\pm
\cos\frac{\theta}{2}
\sin\frac{\theta}{2}
e^{-i\phi}X_{AB}.
\label{Cpm}
\end{equation}
The corresponding measurement probabilities are
\begin{align}
p_+
&=
\cos^2\frac{\theta}{2}(1-X_{AA})
+
\sin^2\frac{\theta}{2}X_{AA},
\nonumber\\
p_-
&=
\sin^2\frac{\theta}{2}(1-X_{AA})
+
\cos^2\frac{\theta}{2}X_{AA}.
\label{measurement-probabilities}
\end{align}
It follows directly that $p_++p_-=1$.

The two eigenvalues of each conditional state are
\begin{equation}
\lambda_{s,\pm}
=
\frac{1\pm\Delta_s}{2},
\qquad
\Delta_s
=
\frac{
\sqrt{(A_s-B_s)^2+4|C_s|^2}
}{p_s}.
\label{conditional-eigenvalues}
\end{equation}
Consequently,
\begin{equation}
S(\rho_{B|s})
=
H\left(\frac{1+\Delta_s}{2}\right),
\label{entropy}
\end{equation}
and the conditional entropy associated with the measurement direction
$\theta$ is
\begin{equation}
S(B|\theta)
=
\sum_{s=\pm}
p_s
H\left(\frac{1+\Delta_s}{2}\right).
\label{conditional-entropy-theta}
\end{equation}
The phase $\phi$ does not affect the conditional entropy because
\begin{equation}
|C_s|
=
\cos\frac{\theta}{2}
\sin\frac{\theta}{2}
|X_{AB}|.
\end{equation}

Rather than assuming that the optimal measurement necessarily occurs
only at $\theta=0$ or $\theta=\pi/2$, we minimize the conditional
entropy over the complete independent range of measurement directions:
\begin{equation}
S_{\min}(B|A)
=
\min_{0\leq\theta\leq\pi/2}
\left[
\sum_{s=\pm}
p_s
H\left(\frac{1+\Delta_s}{2}\right)
\right].
\label{SB-A}
\end{equation}

The interval may be restricted to
$0\leq\theta\leq\pi/2$
because the conditional entropy is invariant
under $\theta\rightarrow\pi-\theta$,
which exchanges the measurement outcomes
up to irrelevant phase factors.

The classical correlation obtained by measuring detector $A$
is therefore
\begin{equation}
J(B|A)
=
H(X_{BB})-S_{\min}(B|A).
\label{JAB}
\end{equation}
Finally, substituting ~\ref{IAB1} and
\ref{JAB} into~\ref{discord},
we obtain
\begin{equation}
\mathcal{D}(B|A)
=
H(X_{AA})
+
\sum_{i=1,3,4}\alpha_i\log\alpha_i
+
S_{\min}(B|A).
\label{Discord}
\end{equation}


For the positive approximate state
$\widetilde{\rho}_{AB}^{(2)}$,
the resulting correlations satisfy
\begin{equation}
0\leq\mathcal{D}(B|A)\leq I(A:B).
\label{discord-bound}
\end{equation}

In our numerical calculations, we verify this
relation together with the positivity condition
\begin{equation}
|X_{AB}|^2
\leq
X_{AA}X_{BB}.
\label{positivity-condition}
\end{equation}
These conditions are satisfied throughout the
parameter region displayed in the figures.

The above expressions characterize the total and quantum correlations
of the positive approximate reduced state
constructed from the second-order detector
response functions.
In the following section, we evaluate these quantities
for a scalar field confined inside a cylindrical cavity.

\section{Analysis of Quantum Correlations in Cavity}
\label{Section4}
We now apply this formalism to the case of a massless scalar field confined within a cylindrical cavity. The presence of the cavity modifies the field correlations through the discrete mode structure imposed by the boundary conditions. Our goal is to evaluate the response functions explicitly and analyze how the cavity affects entanglement, total correlations, and quantum discord between the detectors. 
\begin{figure}
    \centering    \includegraphics[width=1.0\linewidth]{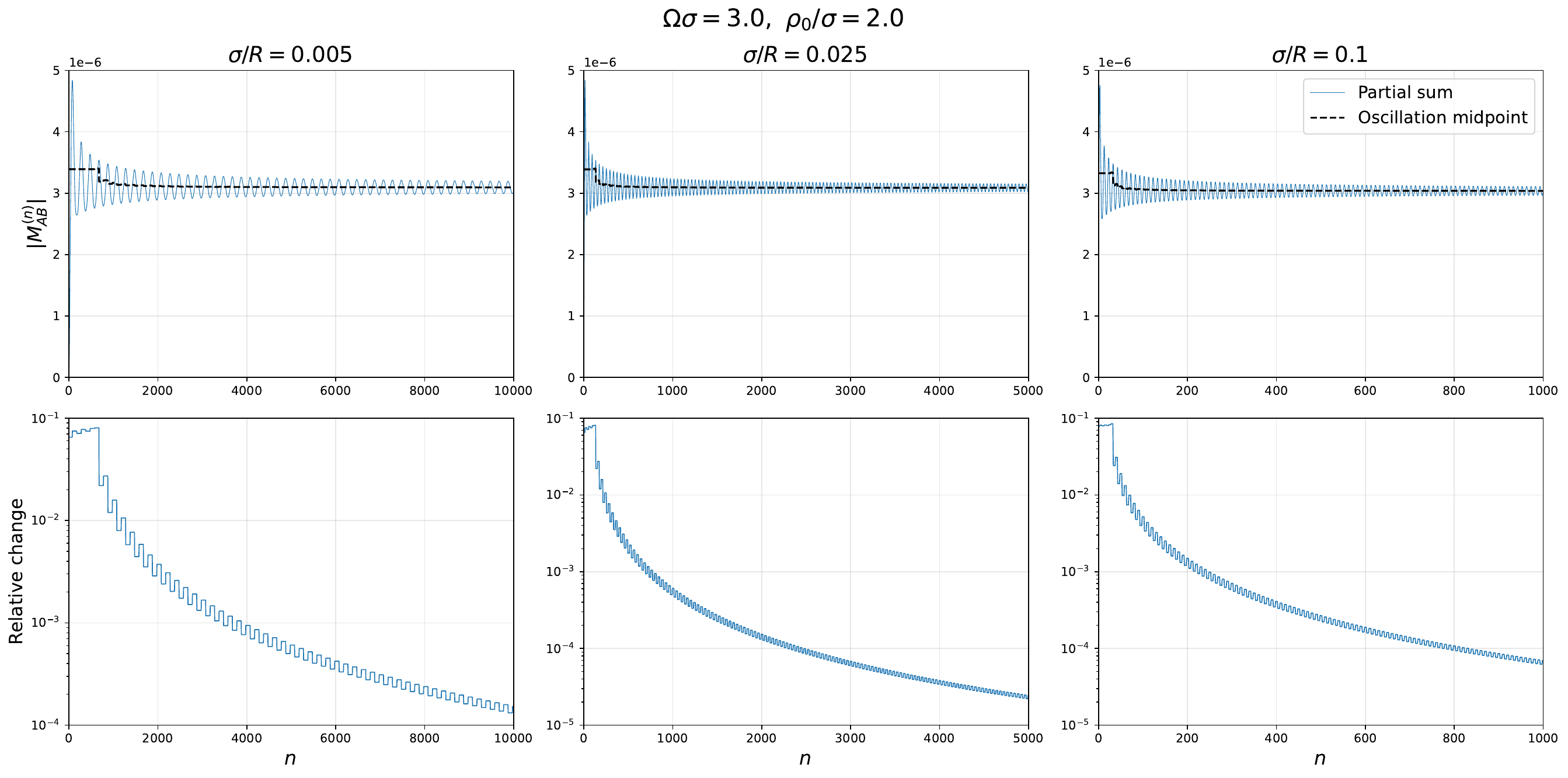}
    \includegraphics[width=1.0\linewidth]{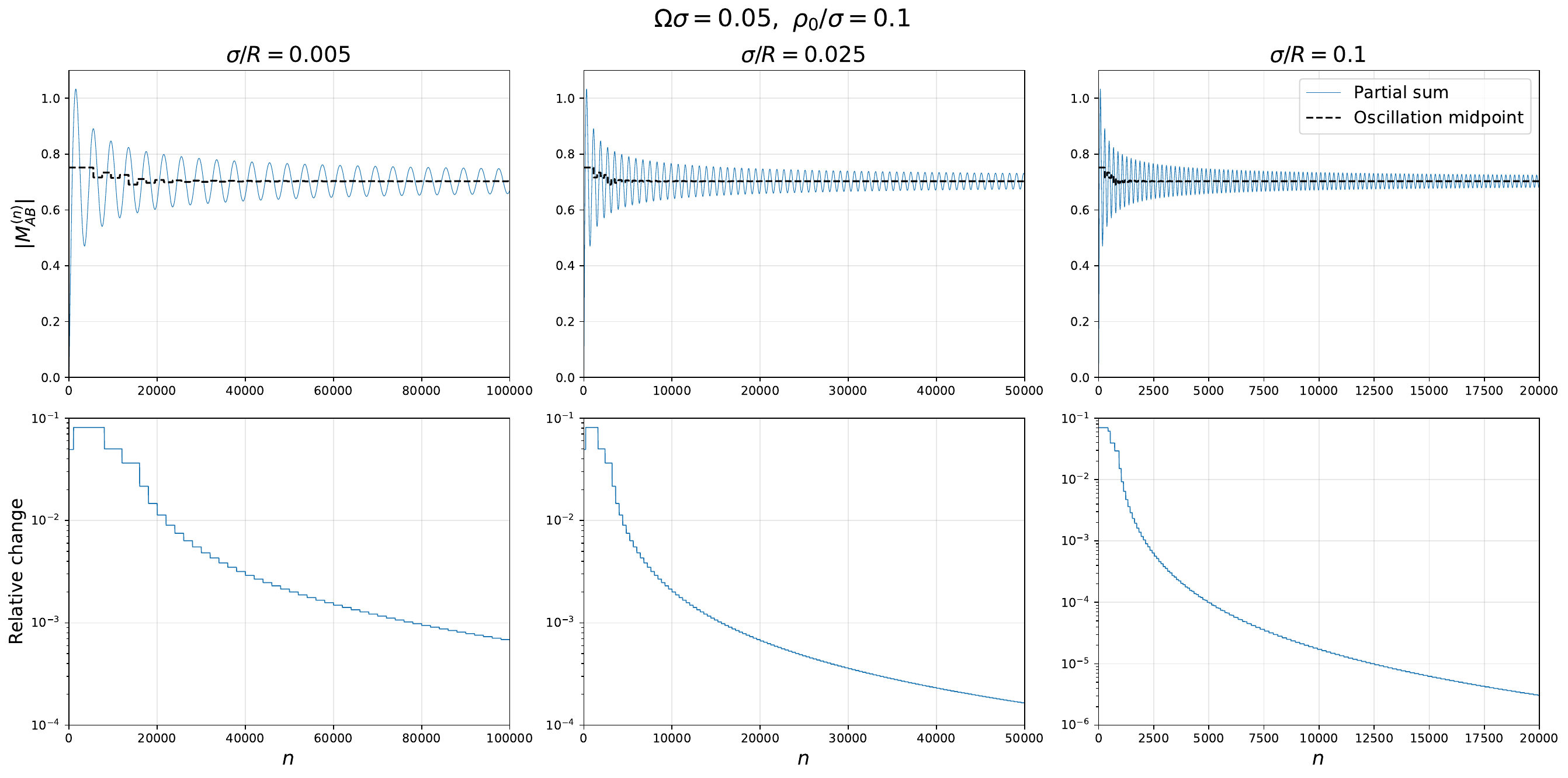}    
  \caption{\it{\small Partial sums of the nonlocal correlation term
$|M_{AB}^{(n)}|/\lambda^2$ as functions of the radial-mode cutoff
$n$ for different values of $\sigma/R$. The upper panel in each
two-row block shows the oscillatory partial sums together with the
oscillation-midpoint estimate (black dashed curve), while the lower
panel shows the relative variation of the midpoint estimate over an adaptive cutoff interval $\Delta n$ determined by the oscillation frequency. The top two rows correspond to $\Omega\sigma=0.05$ and
$\rho_0/\sigma=0.1$, whereas the bottom two rows correspond to
$\Omega\sigma=3.0$ and $\rho_0/\sigma=2.0$.} The results demonstrate 
the oscillatory behavior of the nonlocal mode sum
and the numerical stabilization of its
oscillation-midpoint estimate, with stabilization
generally becoming faster as $\sigma/R$ increases
for the parameter values considered.
}
    \label{fig:1-ParSum}
\end{figure}
Substituting the Wightman functions derived in \cref{subsec:wightman} in \ref{Xij+-'} and \ref{Mij+-'}, we obtain the detector response functions as
\begin{align}
     \label{Xaa11}   X_{AA}&=\frac{\lambda^2\sigma^2}{2\pi R^2}\sum_{n}\frac{1}{J_1^2(\xi_{0n})}\int_{-\infty}^{\infty}ds \;e^{-\sigma^2\big(\Omega+\frac{\xi_{0n}}{R}\cosh{s}\big)^2},\\
         \label{Xbb11}        X_{BB}
&=
\frac{\lambda^2\sigma^2}{2\pi R^2}
\sum_{m,n}
\frac{(J_m(\xi_{mn}\rho_0/R))^2}{J_{|m|+1}^2(\xi_{mn})}
\int_{-\infty}^{\infty}ds\;
e^{-\sigma^2\left(\Omega+\frac{\xi_{mn}}{R}\cosh{s}\right)^2},\\
  \label{XAB}
    X_{AB}&=\frac{\lambda^2\sigma^2}{2\pi R^2}
\sum_{n}
\frac{J_0(\xi_{0n}\rho_0/R)}{J_{1}^2(\xi_{0n})}
\int_{-\infty}^{\infty}ds\;
e^{-\sigma^2\left(\Omega+\frac{\xi_{0n}}{R}\cosh{s}\right)^2},\\
 \label{mab2}
M_{AB}&=\frac{\lambda^2\sigma^2}{2\pi R^2}e^{-\Omega^2\sigma^2}\sum_ne^{-\xi_{0n}^2\sigma^2/(2R^2)}\frac{J_0(\xi_{0n}\rho_0/R)}{J_1^2(\xi_{0n})}\Big(-K_0\Big(\frac{\xi_{0n}^2\sigma^2}{2R^2}\Big)+i\pi I_0\Big(\frac{\xi_{0n}^2\sigma^2}{2R^2}\Big)\Big).
\end{align}

Details of the intermediate steps leading to the above
expressions are provided in \ref{A1}. The mode sums entering the local response terms
$X_{AA}$ and $X_{BB}$ and the cross-correlation
term $X_{AB}$ are numerically convergent, with
stabilization typically achieved using cutoffs
of order $n\sim10^3$ and $m\sim10$.
By contrast, the nonlocal term $M_{AB}$
exhibits slowly decaying oscillations in its
radial-mode partial sums.
As illustrated in~\ref{fig:1-ParSum},
the oscillation-midpoint estimate approaches
a numerically stable value, although the
partial sums themselves exhibit persistent
oscillatory behavior.
We estimate the nonlocal contribution by taking
the midpoint between successive extrema
of the oscillatory partial sums.
The slow convergence reflects the sensitivity
of pointlike detectors to high-frequency cavity
modes. Introducing spatially smeared detector
profiles to account for a finite detector size
can improve the ultraviolet behavior and accelerate
the numerical convergence, as considered in
~\cite{Strohle:2026gdr}. We express our results in terms of the dimensionless parameters
$\Omega\sigma$, $\rho_0/\sigma$, and $\sigma/R$, where $\sigma/R$
controls the cavity radius relative to the interaction timescale.

\begin{figure}
    \centering
          \includegraphics[width=1\linewidth]{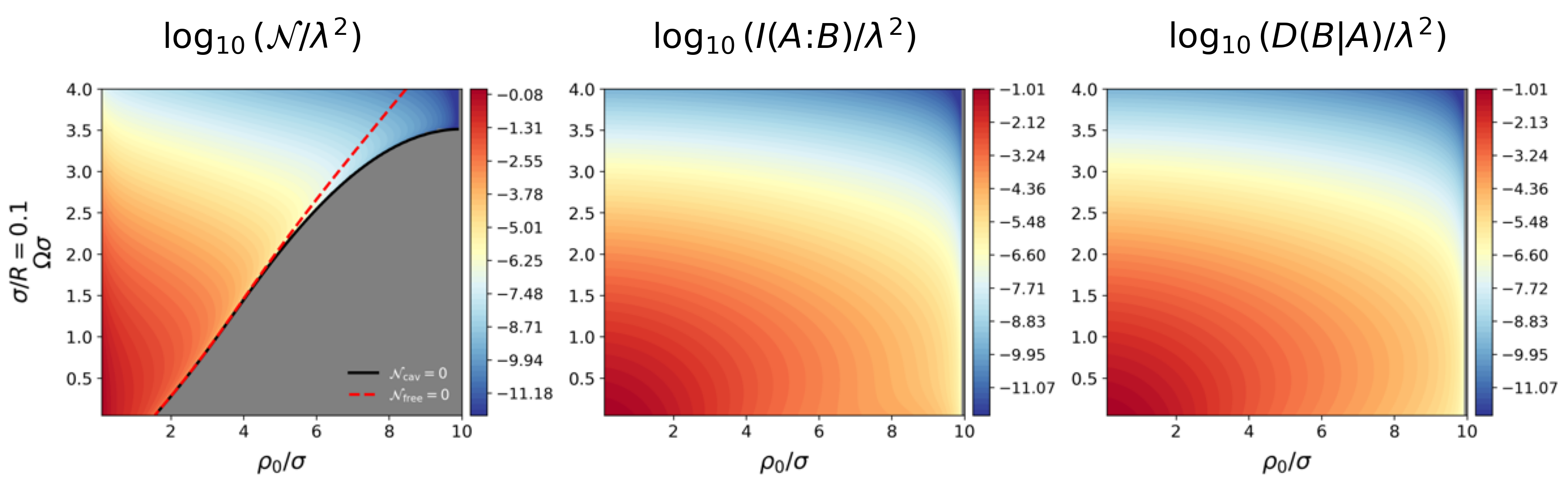}
          \caption{\it{\small Density plots of the entanglement negativity
$\mathcal{N}$, quantum mutual information
$I(A:B)$, and quantum discord $D(B|A)$,
from left to right, as functions of
$\rho_0/\sigma$ and $\Omega\sigma$
for $\sigma/R=0.1$.
All quantities are normalized by $\lambda^2$
and displayed on a base-10 logarithmic scale.
In the negativity panel, the
black solid and red dashed curves denote the
cavity and free-space entanglement thresholds,
respectively. The grey region represents
vanishing cavity negativity.
For $\sigma/R=0.1$, the cavity entangled region
extends beyond the free-space threshold in part
of the displayed range at larger energy gaps,
as further illustrated in~\ref{fig:4-1dSlice}.
Mutual information and quantum discord remain
nonzero over a broader parameter range,
including regions where the negativity vanishes.
For $\sigma/R=0.1$, both entropic quantities
are increasingly suppressed as
$\rho_0/R=(\rho_0/\sigma)(\sigma/R)$ approaches
unity, corresponding to detector B approaching
the cavity wall.}}
    \label{fig:3-Contour}
\end{figure}

\begin{figure}
    \centering
    \includegraphics[width=1.0\linewidth]{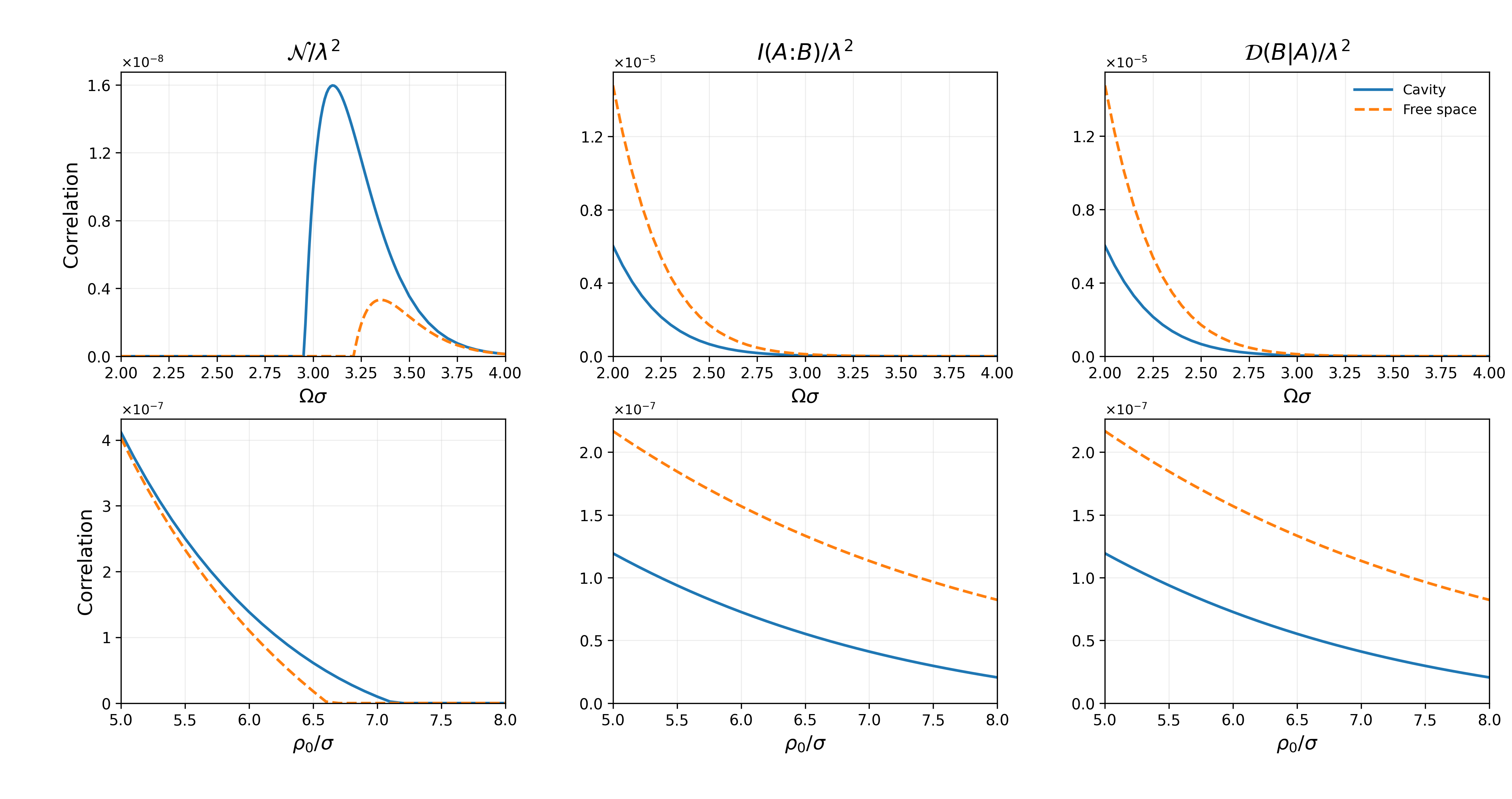}     
\caption{\it{\small
Harvested correlations in one-dimensional slices
of the parameter space. The columns show the entanglement negativity
$\mathcal{N}$, quantum mutual information
$I(A:B)$, and quantum discord
$\mathcal{D}(B|A)$, respectively.
The upper row shows their dependence on the
dimensionless energy gap $\Omega\sigma$ at fixed
$\rho_0/\sigma=7$ and $\sigma/R=0.1$, while
the lower row shows their dependence on the
normalized detector separation $\rho_0/\sigma$
at fixed $\Omega\sigma=3$ and $\sigma/R=0.1$.
All quantities are normalized by $\lambda^2$.
The blue solid and orange dashed curves
represent the cavity and free-space results,
respectively.
For the selected parameters, the cavity supports
nonzero negativity over a broader parameter
range than free space. In contrast, both mutual
information and quantum discord remain larger
in free space throughout the displayed parameter
ranges. These results demonstrate that an
enhancement of entanglement negativity does not
necessarily imply an enhancement of total
or discord-type quantum correlations.
}}
    \label{fig:4-1dSlice}
\end{figure}

The normalized negativity corresponds
to its leading-order contribution, whereas
mutual information and quantum discord are
evaluated by directly substituting the normalized
response coefficients into the positive approximate
state $\widetilde{\rho}_{AB}^{(2)}$, corresponding
to an effective choice $\lambda=1$.
This convention is used consistently in the
numerical evaluation of the entropic quantities
presented in~\ref{fig:3-Contour} and
\ref{fig:4-1dSlice}.
The resulting entropic quantities contain
higher-order contributions associated with
this approximation. Checks at representative
parameter values with smaller coupling strengths
confirm that their normalized values approach
the same leading-order limit.

The
parameter region is restricted by
\begin{equation}
\frac{\rho_0}{\sigma}\frac{\sigma}{R}
=
\frac{\rho_0}{R}<1,
\end{equation}
which ensures that the off-axis detector remains inside the cavity. The convergence of the nonlocal mode sum depends
sensitively on these parameters. For sufficiently
large values of $\Omega\sigma$ and $\rho_0/\sigma$,
numerical stability is typically achieved
with a mode cutoff of order $n\sim10^3$.
In contrast, the small-$\sigma/R$ regime converges
considerably more slowly and require cutoffs
as large as $n\sim10^5$, as illustrated
in~\ref{fig:1-ParSum}. The cutoff is chosen such that
the relative change is less than  $10^{-3}$ for the most
numerically demanding parameter values considered.
This criterion characterizes the stabilization
of the oscillation-midpoint estimate rather than
providing a rigorous bound on the truncation
error of the infinite mode sum. ~\ref{fig:1-ParSum} shows that the
stabilized value of the nonlocal contribution
varies only weakly with $\sigma/R$ over much
of the parameter range considered.

\begin{figure}
    \centering
        \includegraphics[width=0.66\linewidth]{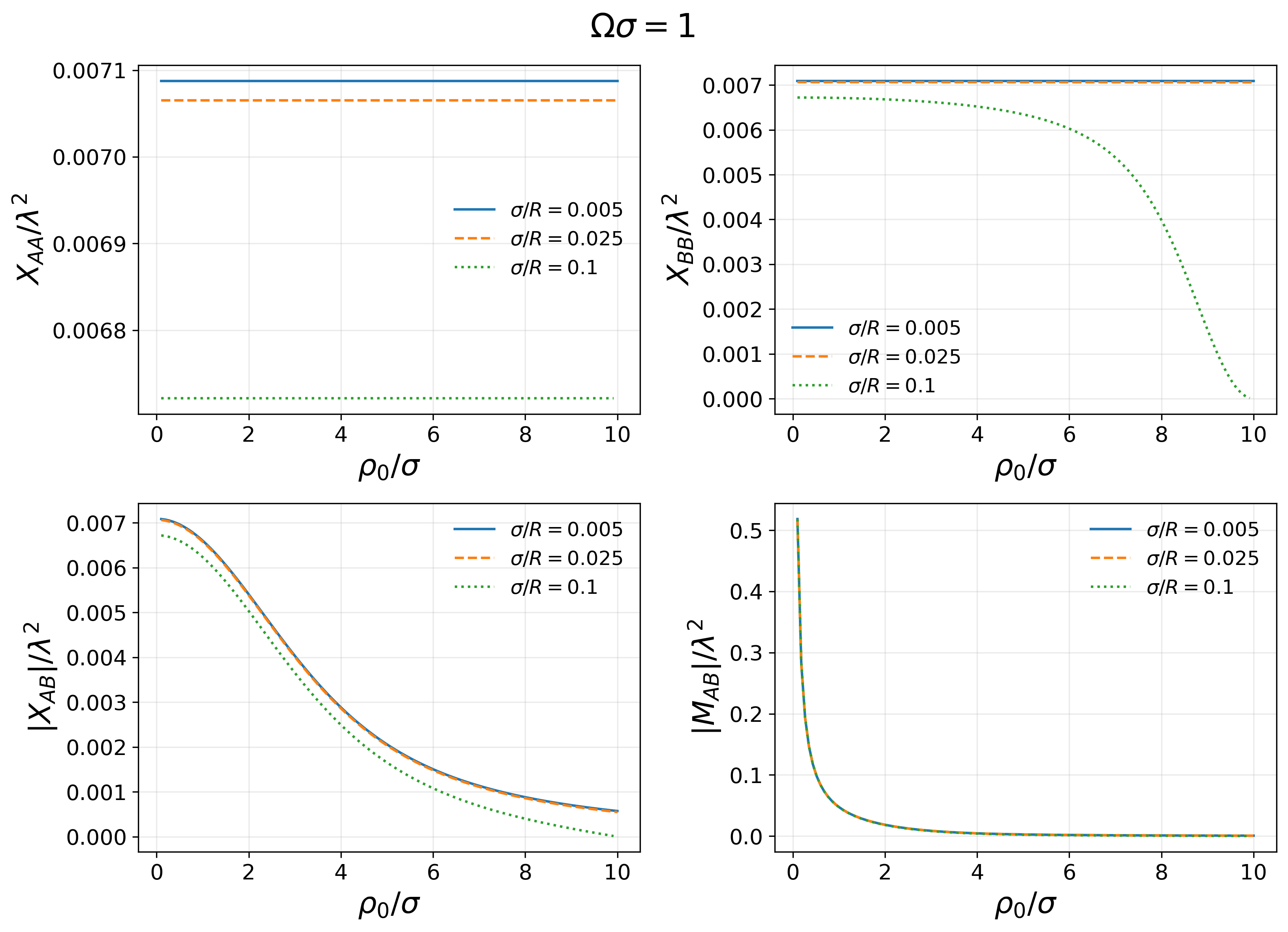}
      
\caption{
\it{\small Behavior of the normalized detector-correlation
terms $X_{AA}/\lambda^2$, $X_{BB}/\lambda^2$,
$|X_{AB}|/\lambda^2$, and $|M_{AB}|/\lambda^2$
as functions of the normalized detector separation
$\rho_0/\sigma$ for different values of
$\sigma/R$ at fixed $\Omega\sigma=1$.
Since detector A remains on the cavity axis,
$X_{AA}/\lambda^2$ is independent of the
detector separation. The off-axis response $X_{BB}/\lambda^2$ and
single-excitation coherence $|X_{AB}|/\lambda^2$
are suppressed as detector B approaches the
Dirichlet boundary, while $|M_{AB}|/\lambda^2$
decreases rapidly with separation and exhibits
a relatively weak dependence on $\sigma/R$.}
}
\label{fig:2-Corr}
\end{figure}

The dependence of the individual response
and correlation functions on the detector
separation is shown in~\ref{fig:2-Corr}.
The on-axis local response $X_{AA}$ is
independent of $\rho_0/\sigma$, whereas
$X_{BB}$, $X_{AB}$, and $M_{AB}$ depend
on the position of the off-axis detector
and on the cavity parameter.
In particular,
the off-axis local response and
cross-correlation terms are suppressed
as detector B approaches the Dirichlet wall.

Using ~\ref{NegApprox}, \ref{IAB1}, and \ref{Discord}, we numerically evaluate the
entanglement negativity, quantum mutual information, and quantum discord. \ref{fig:3-Contour} shows density
plots of these quantities as functions of the normalized
detector separation $\rho_0/\sigma$ and the dimensionless
energy gap $\Omega\sigma$ for fixed $\sigma/R$.
In the negativity panel, the black solid
curve denotes the cavity entanglement threshold
$\mathcal{N}_{\mathrm{cav}}=0$, while the red dashed curve
denotes the corresponding free-space threshold
$\mathcal{N}_{\mathrm{free}}=0$. We see, the detectors has finite entanglement negativity for larger energy gaps as the off-axis detector reach the cavity boundary, as illustrated by the one-dimensional comparisons
in~\ref{fig:4-1dSlice}. Notably, free space entanglement negativity is zero for this parameter regime. Thus, cavity boundaries assist with the entanglement harvesting for large energy gaps, though the magnitude of negativity is small in this regime.

In contrast, the mutual information and quantum discord
remain nonzero over a broader range of
detector separations and energy gaps, including regions
where the negativity has vanished. Both quantities
decrease smoothly with increasing energy gap and
detector separation. As shown in
\ref{fig:3-Contour}, their suppression becomes particularly
pronounced as
$(\rho_0/\sigma)(\sigma/R)=\rho_0/R$
approaches unity, corresponding to the off-axis
detector approaching the cavity wall. The numerical
results satisfy
$0 \leq D(B|A) \leq I(A:B)$
throughout the investigated parameter region.

To examine these trends more closely, we present one-dimensional
sections of the parameter space in~\ref{fig:4-1dSlice}.
The first row shows the dependence of the three correlation measures
on $\Omega\sigma$ at fixed $\rho_0/\sigma=7$ and $\sigma/R=0.1$,
while the second row shows their dependence on $\rho_0/\sigma$
at fixed $\Omega\sigma=3$ and $\sigma/R=0.1$. The columns correspond to
the entanglement negativity, quantum mutual information,
and quantum discord, respectively. The solid blue and dashed orange curves represent the
cavity and free-space results, respectively. The parameter regime considered here corresponds to the case where the free-space correlations are most different from the cavity correlations. The behavior for other representative parameter choices is presented in \ref{app:extended-1d-comparisons}.


Interestingly,~\ref{fig:4-1dSlice} reveals
a distinct response of the three correlation
measures to cavity confinement. The cavity supports nonzero
entanglement negativity over a broader separation
than free space. Moreover, the negativity gets enhanced as compared to free space negativity for larger energy gaps as the off-axis detector reaches the boundary. On the other hand, both the mutual
information and quantum discord remain smaller
than their free-space counterparts throughout
the investigated parameter regimes. This indicates
that cavity confinement can favor entanglement
harvesting without necessarily enhancing total
or discord-type quantum correlations. Same trends are observed for other parameter regions, see free-space vs cavity behavior of quantum correlations in~\ref{app:extended-1d-comparisons}.

The behavior of the observables depicted above can be better understood by looking at the behavior of individual detector-correlation
terms, illustrated in~\ref{fig:2-Corr}, where
$X_{AA}$, $X_{BB}$, $|X_{AB}|$, and $|M_{AB}|$
are plotted as functions of $\rho_0/\sigma$.
Because detector $A$ remains fixed on the symmetry
axis, its local excitation probability $X_{AA}$
is independent of the detector separation.
By contrast, $X_{BB}$ varies with the position
of detector $B$ and approaches zero as the
detector approaches the cavity wall, consistently
with the Dirichlet boundary condition.
The cross-correlation $|X_{AB}|$ decreases
with detector separation over the displayed
range and vanishes as detector $B$ approaches
the Dirichlet boundary.
The nonlocal term $|M_{AB}|$ decreases rapidly
with detector separation and exhibits only a
weak dependence on $\sigma/R$ over the range
considered.

The different dependence of these correlation
terms explain why the negativity
can vanish beyond a finite detector separation,
whereas the mutual information and quantum discord
remain nonzero over a broader parameter range.
The negativity is determined by the competition
between the nonlocal correlation $|M_{AB}|$
and the local detector noise. In contrast,
the mutual information and quantum discord depend
on the positive approximate
reduced two-detector state, including the
cross-correlation $X_{AB}$, and can therefore
remain nonzero even after the condition for
entanglement harvesting is no longer satisfied.
More generally, the persistence of quantum discord
after the disappearance of entanglement has also
been observed in relativistic settings
\cite{Datta:2009wug}.
In the present cavity configuration, however,
both the mutual information and quantum discord
are smoothly suppressed as the detector separation
or energy gap increases
and exhibit stronger suppression as the off-axis
detector approaches the Dirichlet boundary.

\section{Conclusion}\label{Section5}

In this work, we investigate the effects of cavity boundaries
on harvested correlations between two Unruh--DeWitt detectors
coupled to a scalar field inside a cylindrical cavity.
Going beyond entanglement as the sole measure of quantum
correlations, we characterize the harvested correlations
using negativity, mutual information, and quantum discord
for stationary detectors in a cavity. For smaller energy gaps and for small separations, the entanglement negativity follows the free space behavior. For larger energy gap, the entanglement negativity gets enhanced as the off-axis  detector reach cavity boundary and negativity is nonzero for separations where the free space negativity is zero.
Both mutual information and quantum discord decrease
smoothly with detector separation and are increasingly
suppressed as the off-axis detector approaches the cavity
boundary.

The origin of this separation lies in the distinct imprint
of the cavity-modified mode structure on the detector
correlation terms. The nonlocal correlation contribution
$M_{AB}$, which governs entanglement harvesting, is
strongly suppressed with increasing detector separation,
whereas the cross-correlation term $X_{AB}$ remains
comparatively robust over a broader range.
Simultaneously, the local noise term $X_{BB}$ is
suppressed near the boundary, highlighting the
asymmetry between the detectors arising from boundary
conditions that break translational invariance.
Together, these effects generate regimes in which the leading-order entanglement negativity vanishes,
while more general quantum correlations persist.
Throughout the parameter region investigated, the
resulting correlations satisfy the expected hierarchy
$0\leq D(B|A)\leq I(A:B)$. 

For the parameters shown in~\ref{fig:4-1dSlice},
cavity confinement can extend the region of nonzero
entanglement negativity beyond the corresponding
free-space threshold, while the mutual information and quantum discord evaluated
from the positive approximate state remain smaller
than their free-space counterparts.
This demonstrates that conditions favoring entanglement
harvesting do not necessarily enhance other measures
of quantum correlations. Cylindrical cavities therefore provide a controlled
setting to probe the distinction between entanglement
and broader nonclassical correlations.
Our results may also be relevant for cavity-based
analogue quantum simulation platforms and studies
of relativistic quantum information in bounded
geometries \cite{Haroche2006,Walther2006}.

The present calculation describes an infinitely
long cylindrical cavity with a continuous
longitudinal mode spectrum.
Its geometric free-space limit is therefore
obtained by taking $R\rightarrow\infty$ at fixed
detector separation, switching duration, and
energy gap. For a finite-length cavity, recovery
of free-space behavior would additionally require
$L\rightarrow\infty$. Extending the present analysis to a finite-length
cavity may introduce additional numerical
convergence challenges associated with the
nonlocal correlation term. One possible resolution is to consider smeared
field operators to incorporate
the finite size of the detectors,
as considered in~\cite{Strohle:2026gdr}. Such a treatment provides a physically motivated
regularization and may improve the ultraviolet
behavior of the detector--field interaction
\cite{Martin-Martinez:2012ysv,Pozas-Kerstjens:2015gta}.
A systematic investigation of finite-length
cavities and different detector-smearing
profiles is therefore left for future work.

A natural extension of the present work is to
investigate accelerated detector configurations
and boundary-assisted harvesting protocols, where
acceleration, detector geometry, and reflecting
boundaries are known to strongly influence
entanglement harvesting
\cite{Liu:2020jaj, Li:2024dvs,Barman:2023wkr,Ma:2024tyn,
Wu:2026woq,Salomaa:2026xzn,Huang:2025gme}.
Previous studies in these settings have focused
primarily on entanglement generation and coherence
dynamics. In light of the hierarchy observed in
the present analysis, it would be worthwhile
to examine whether more general quantum
correlations, such as quantum discord and
mutual information, remain robust in regimes
where entanglement is suppressed by acceleration,
environmental effects, or
boundary-induced modifications of the field
correlations.
Such investigations could provide deeper insight
into the interplay between
detector motion,
spacetime geometry,
boundary conditions,
and the persistence of nonclassical correlations
in relativistic quantum systems.

\section*{Acknowledgments}
We thank Kinjalk Lochan for helpful comments and suggestions. This work was financially supported by Vellore Institute of Technology (VIT), Vellore under the Faculty Seed Grant (RGEMS) (Sanction Order No.: SG20260125). HSS thanks Indian Institute of Technology, Madras for support through postdoctoral fellowship.

\appendix
\labelformat{section}{Appendix #1}

\section{Derivation of pullback of Wightman function on stationary trajectory in cavity}
\label{app:kz_integral}
For a detector at rest inside the cylindrical cavity, the Wightman function takes the form
\begin{align}
\mathcal{W}(\tau_-)
=\frac{1}{(2\pi R)^2}
\sum_{m,n}
\frac{J_m^2(\xi_{mn}\eta)}{J_{|m|+1}^2(\xi_{mn})}
\int_{-\infty}^{\infty}\frac{dk_z}{\omega_k}
e^{-i\omega_k\tau_-},
\label{WtauIn}
\end{align}
where
\[
\omega_k=\sqrt{k_z^2+\frac{\xi_{mn}^2}{R^2}}.
\]

We define the longitudinal momentum integral
\begin{equation}
I(\tau_-)
=\int_{-\infty}^{\infty}\frac{dk_z}{\omega_k}
e^{-i\omega_k\tau_-}
=\int_{-\infty}^{\infty}
\frac{dk_z}{\sqrt{k_z^2+\frac{\xi_{mn}^2}{R^2}}}
e^{-i\sqrt{k_z^2+\frac{\xi_{mn}^2}{R^2}}\tau_-}.
\label{I}
\end{equation}

Since the integrand is even in $k_z$, the integral may be written as
\begin{equation}
I(\tau_-)
=2\int_{0}^{\infty}
\frac{dk_z}{\sqrt{k_z^2+\frac{\xi_{mn}^2}{R^2}}}
e^{-i\sqrt{k_z^2+\frac{\xi_{mn}^2}{R^2}}\tau_-},
\qquad (\tau_->0).
\end{equation}

Introducing
\begin{equation}
I'(\tau_-)
=\int_{0}^{\infty}
\frac{dk_z}{\sqrt{k_z^2+\frac{\xi_{mn}^2}{R^2}}}
e^{-i\sqrt{k_z^2+\frac{\xi_{mn}^2}{R^2}}\tau_-},
\qquad \tau_->0,
\end{equation}
and performing the substitution
\[
k_z=\frac{\xi_{mn}}{R}\sinh u ,
\]
for which
\[
\sqrt{k_z^2+\frac{\xi_{mn}^2}{R^2}}=\frac{\xi_{mn}}{R}\cosh u,
\qquad
dk_z=\frac{\xi_{mn}}{R}\cosh u\,du,
\]
the integral simplifies to
\begin{equation}
I'(\tau_-)
=\int_{0}^{\infty}du\,
e^{-i\frac{\xi_{mn}}{R}\tau_-\cosh u}.
\end{equation}

This expression is the analytic continuation of the standard integral representation of the modified Bessel function
\begin{equation}
K_0(z)
=\int_{0}^{\infty}du\,e^{-z\cosh u},
\qquad \Re(z)>0.
\end{equation}

For the Wightman function, convergence is ensured by the usual
$i\epsilon$ prescription,
\[
\tau_- \rightarrow \tau_- - i\epsilon,
\qquad \epsilon>0.
\]
Taking the limit $\epsilon\to0^+$ gives
\begin{equation}
I'(\tau_-)
=K_0\Big(i\frac{\xi_{mn}}{R}\tau_-\Big)
=-i\frac{\pi }{2}
H_0^{(2)}\Big(\frac{\xi_{mn}}{R}\tau_-\Big),
\qquad \tau_->0,
\end{equation}
where $H_0^{(2)}$ denotes the Hankel function of the second kind.

Substituting this result into ~\ref{I}, we obtain
\begin{equation}
I(\tau_-)
=-i\pi \,H_0^{(2)}(\frac{\xi_{mn}}{R}\tau_-),
\qquad \tau_->0.
\label{Ipos}
\end{equation}

For arbitrary real time separation $-\infty<\tau_-<\infty$, the
$i\epsilon$ prescription fixes the analytic structure uniquely.
Using the identities \cite{gradshteyn2007} 
\begin{align}
\int_{-\infty}^{\infty}dx\,e^{-iz\cosh x}
&=-i\pi H_0^{(2)}(z), \\
\int_{-\infty}^{\infty}dx\,e^{iz\cosh x}
&= i\pi H_0^{(1)}(z),
\qquad z>0,
\end{align}
we obtain
\begin{equation}
I(\tau_-)
=-i\pi\left[
\theta(\tau_-)H_0^{(2)}\Big(\frac{\xi_{mn}}{R}\tau_-\Big)
-\theta(-\tau_-)H_0^{(1)}\Big(-\frac{\xi_{mn}}{R}\tau_-\Big)
\right].
\label{Ifull}
\end{equation}

Substituting \ref{Ifull} into \ref{WtauIn}, the Wightman function becomes 
\begin{align}
\mathcal{W}(\tau_-)
=
-\frac{i}{4\pi R^2}
\sum_{m,n}
\frac{J_m^2(\xi_{mn}\eta)}{J_{|m|+1}^2(\xi_{mn})}
\left[
\theta(\tau_-)H_0^{(2)}\Big(\frac{\xi_{mn}}{R}\tau_-\Big)
-\theta(-\tau_-)H_0^{(1)}\Big(-\frac{\xi_{mn}}{R}\tau_-\Big)
\right].
\end{align}

\section{Derivation of elements of reduced density matrix}
\label{A1}
Substituting \ref{WAA} into \ref{Xij+-'}, we have
\begin{equation}
\begin{split}
\label{XAApm}
X_{AA}= -\frac{i \sigma \lambda^2}{4\sqrt{\pi} R^2}
\sum_{n}\frac{1}{J_1^2(\xi_{0n})}
&\int_{-\infty}^{\infty} d\tau_-^{AA}\;
e^{-(\tau_-^{AA})^2/(4\sigma^2)}
e^{-i\Omega\tau_-^{AA}}\\
&\left[
\theta(\tau_-^{AA})H_0^{(2)}\Big(\frac{\xi_{0n}}{R}\tau_-^{AA}\Big)
-\theta(-\tau_-^{AA})H_0^{(1)}\Big(-\frac{\xi_{0n}}{R}\tau_-^{AA}\Big)
\right].
\end{split}
\end{equation}
Using the definition of the Heaviside theta function, the integral over
$\tau_-^{AA}$ splits into positive and negative domains. After changing
$\tau_-^{AA}\rightarrow -\tau_-^{AA}$ in the second integral, we obtain
\begin{equation}
\begin{split}
\label{XAApos}
X_{AA}=
-\frac{i\sigma\lambda^2}{4\sqrt{\pi}R^2}
\sum_{n}\frac{1}{J_1^2(\xi_{0n})}
\int_{0}^{\infty} d\tau_-^{AA}\;
e^{-(\tau_-^{AA})^2/(4\sigma^2)}
\Big[
e^{-i\Omega\tau_-^{AA}} H_0^{(2)}\Big(\frac{\xi_{0n}}{R}\tau_-^{AA}\Big)
-
e^{i\Omega\tau_-^{AA}} H_0^{(1)}\Big(\frac{\xi_{0n}}{R}\tau_-^{AA}\Big)
\Big].
\end{split}
\end{equation}
Using $H^{(1)}_0(z)=[H^{(2)}_0(z)]^*$, the expression can be written in terms of the imaginary part as follows
\begin{equation}
    \label{ImH}
    \begin{split}
    X_{AA} =-\frac{\sigma\lambda^2}{2\sqrt{\pi} R^2}\sum_n\frac{1}{J_1^2(\xi_{0n})}\text{Im}\Bigg(\int_{0}^{\infty}d\tau^{AA}_-e^{-(\tau^{AA}_-)^2/4\sigma^2}e^{-i\Omega\tau_-^{AA}}H_0^{(2)}\Big(\frac{\xi_{0n}}{R}\tau_-^{AA}\Big)\Bigg).
    \end{split}
\end{equation}

We now consider the integral
\begin{equation}
\begin{split}
I &=
\int_{0}^{\infty}
d\tau_-^{AA}
\,e^{-(\tau_-^{AA})^2/4\sigma^2}
e^{-i\Omega\tau_-^{AA}}
H_0^{(2)}\!\Big(\frac{\xi_{0n}}{R}\tau_-^{AA}\Big).
\end{split}
\end{equation}

Using the integral representation of the Hankel function, this integral can be written as
\begin{equation}
\begin{split}
I
&=
\frac{\sigma}{\sqrt{\pi}i}
\int_{-\infty}^{\infty}ds\;
e^{-\sigma^2\left(\Omega+\frac{\xi_{0n}}{R}\cosh{s}\right)^2}
\Bigg(
1+
\mathrm{erf}\Big[
i\sigma\Big(\Omega+\frac{\xi_{0n}}{R}\cosh{s}\Big)
\Big]
\Bigg).
\end{split}
\end{equation}

Using the identity
\[
\mathrm{erf}(ix)=i\,\mathrm{erfi}(x),
\]
the expression becomes
\begin{equation}
\begin{split}
I
&=
-\frac{\sigma i}{\sqrt{\pi}}
\int_{-\infty}^{\infty}ds\;
e^{-\sigma^2\left(\Omega+\frac{\xi_{0n}}{R}\cosh{s}\right)^2}
\Big(
1+i\,\mathrm{erfi}\Big[
\sigma\Big(\Omega+\frac{\xi_{0n}}{R}\cosh{s}\Big)
\Big]
\Big).
\end{split}
\end{equation}

Rearranging the terms yields
\begin{equation}
\begin{split}
I
&=
-\frac{\sigma}{\sqrt{\pi}}
\int_{-\infty}^{\infty}ds\;
e^{-\sigma^2\left(\Omega+\frac{\xi_{0n}}{R}\cosh{s}\right)^2}
\left(
i-\,\mathrm{erfi}\Big[
\sigma\Big(\Omega+\frac{\xi_{0n}}{R}\cosh{s}\Big)
\Big]
\right).
\end{split}
\label{intxaa}
\end{equation}

Substituting ~\ref{intxaa} into .~\ref{ImH} and taking the imaginary part, we obtain
\begin{equation}
\begin{split}
X_{AA}
&=
-\frac{\sigma\lambda^2}{2\sqrt{\pi}R^2}
\sum_n
\frac{1}{J_1^2(\xi_{0n})}
\,\mathrm{Im}
\Bigg(
-\frac{\sigma}{\sqrt{\pi}}
\int_{-\infty}^{\infty}ds\;
e^{-\sigma^2\left(\Omega+\frac{\xi_{0n}}{R}\cosh{s}\right)^2}
\left(
i-\,\mathrm{erfi}\Big[
\sigma\Big(\Omega+\frac{\xi_{0n}}{R}\cosh{s}\Big)
\Big]
\right)
\Bigg).
\end{split}
\end{equation}

Since the $\mathrm{erfi}$ term contributes only to the real part, the imaginary part arises solely from the first term. Consequently, we obtain
\begin{equation}
\begin{split}
X_{AA}
&=
\frac{\lambda^2\sigma^2}{2\pi R^2}
\sum_{n}
\frac{1}{J_1^2(\xi_{0n})}
\int_{-\infty}^{\infty}ds\;
e^{-\sigma^2\left(\Omega+\frac{\xi_{0n}}{R}\cosh{s}\right)^2}.
\end{split}
\end{equation}
Further, substituting \ref{WB} into \ref{Xij+-'} gives
 \begin{equation}
    \label{Xbb1}
    \begin{split}
            X_{BB}&=-\frac{i\sigma \lambda^2}{4\sqrt{\pi}R^2}\sum_{m,n}\frac{(J_m(\xi_{mn}\eta))^2}{J^2_{|m|+1}(\xi_{mn})}\int_{-\infty}^{\infty} d\tau_-^{BB}\;
e^{-(\tau_-^{BB})^2/(4\sigma^2)}
e^{-i\Omega\tau_-^{BB}}\\ &\times\Big[
\theta(\tau_-^{BB})H_0^{(2)}\!\left(\frac{\xi_{mn}}{R}\tau_-^{BB}\right)
-\theta(-\tau_-^{BB})H_0^{(1)}\!\left(-\frac{\xi_{mn}}{R}\tau_-^{BB}\right)
\Big].
                \end{split}
\end{equation}
Following similar steps it can be simplified to
\begin{equation}
\begin{split}
X_{BB}
=
\frac{\lambda^2\sigma^2}{2\pi R^2}
\sum_{m,n}
\frac{(J_m(\xi_{mn}\eta))^2}{J_{|m|+1}^2(\xi_{mn})}
\int_{-\infty}^{\infty}ds\;
e^{-\sigma^2\left(\Omega+\frac{\xi_{mn}}{R}\cosh{s}\right)^2}.
\end{split}
\end{equation}
Next, substituting \ref{WAB} into \ref{Xij+-'}, the cross term $X_{AB}$ between the two detectors can be written as
\begin{equation}
\begin{split}
\label{XABpm}
X_{AB}= -\frac{i \sigma \lambda^2}{4\sqrt{\pi} R^2}
\sum_{n}\frac{J_0(\xi_{0n}\eta)}{J_1^2(\xi_{0n})}
\int_{-\infty}^{\infty} d\tau_-^{AB}\;
e^{-(\tau_-^{AB})^2/(4\sigma^2)}
e^{-i\Omega\tau_-^{AB}}
\left[
\theta(\tau_-^{AB})H_0^{(2)}(\frac{\xi_{0n}}{R}\tau_-^{AB})
-\theta(-\tau_-^{AB})H_0^{(1)}(-\frac{\xi_{0n}}{R}\tau_-^{AB})
\right].
\end{split}
\end{equation}
On simplification, it becomes
\begin{equation}
\begin{split}
        \label{XAB1}
    X_{AB}=\frac{\lambda^2\sigma^2}{2\pi R^2}
\sum_{n}
\frac{J_0(\xi_{0n}\eta)}{J_{1}^2(\xi_{0n})}
\int_{-\infty}^{\infty}ds\;
e^{-\sigma^2\left(\Omega+\frac{\xi_{0n}}{R}\cosh{s}\right)^2}
\end{split}
\end{equation}

The cross term $M_{AB}$ is obtained by substituting \ref{WAB} in \ref{Mij+-'}, yielding
\begin{equation}
\label{Mij+-'''}
\begin{split}
M_{AB} =
\frac{i\sigma\lambda^2}{2\sqrt{\pi}R^2}
e^{-\Omega^2\sigma^2}
\int_{0}^{\infty} d\tau_-^{AB}\;
e^{-(\tau_-^{AB})^2/(4\sigma^2)}
\sum_{n}\frac{J_0(\xi_{0n}\eta)}{J_{1}^2(\xi_{0n})}
\Big[
\theta(\tau_-^{AB})H_0^{(2)}\Big(\frac{\xi_{0n}}{R}\tau_-^{AB}\Big)
-
\theta(-\tau_-^{AB})H_0^{(1)}\Big(-\frac{\xi_{0n}}{R}\tau_-^{AB}\Big)
\Big].
\end{split}
\end{equation}
Since $\tau_-^{AB}\geq0$, only the $H^{(2)}_0$ term contributes and above expression simplifies to
\begin{equation}
\label{Mab1}
\begin{split}
M_{AB}
=
\frac{i\sigma\lambda^2}{2\sqrt{\pi}R^2}
e^{-\Omega^2\sigma^2}
\sum_{n}
\frac{J_0(\xi_{0n}\eta)}{J_{1}^2(\xi_{0n})}
\int_{0}^{\infty}
d\tau_-^{AB}\;
e^{-(\tau_-^{AB})^2/(4\sigma^2)}
H_0^{(2)}\!\Big(\frac{\xi_{0n}}{R}\tau_-^{AB}\Big).
\end{split}
\end{equation}

The remaining integral can be evaluated analytically, yielding
\begin{align}
\label{int1}
\int_{0}^{\infty}
d\tau_-^{AB}\;
e^{-(\tau_-^{AB})^2/(4\sigma^2)}
H_0^{(2)}\!\Big(\frac{\xi_{0n}}{R}\tau_-^{AB}\Big)
=
\frac{\sigma}{\sqrt{\pi}}
e^{-\frac{\xi_{0n}^2\sigma^2}{2R^2}}
\left[
\pi I_0\!\left(\frac{\sigma^2\xi_{0n}^2}{2R^2}\right)
+
i\,K_0\!\left(\frac{\sigma^2\xi_{0n}^2}{2R^2}\right)
\right],
\end{align}
where $I_0$ and $K_0$ denote the modified Bessel functions.

Substituting ~\ref{int1} into ~\ref{Mab1}, we obtain
\begin{equation}
\label{mab11}
\begin{split}
M_{AB}
&=
\frac{\sigma^2\lambda^2}{2\pi R^2}
e^{-\Omega^2\sigma^2}
\sum_{n}
\frac{J_0(\xi_{0n}\eta)}{J_1^2(\xi_{0n})}
e^{-\frac{\xi_{0n}^2\sigma^2}{2R^2}} 
\Bigg[
-K_{0}\!\left(\frac{\sigma^2\xi_{0n}^2}{2R^2}\right)
+
i\pi I_{0}\!\left(\frac{\sigma^2\xi_{0n}^2}{2R^2}\right)
\Bigg].
\end{split}
\end{equation}

\section{Additional parameter-dependent comparisons}
\label{app:extended-1d-comparisons}

\begin{figure*}[t]
    \centering
    \includegraphics[width=\textwidth]{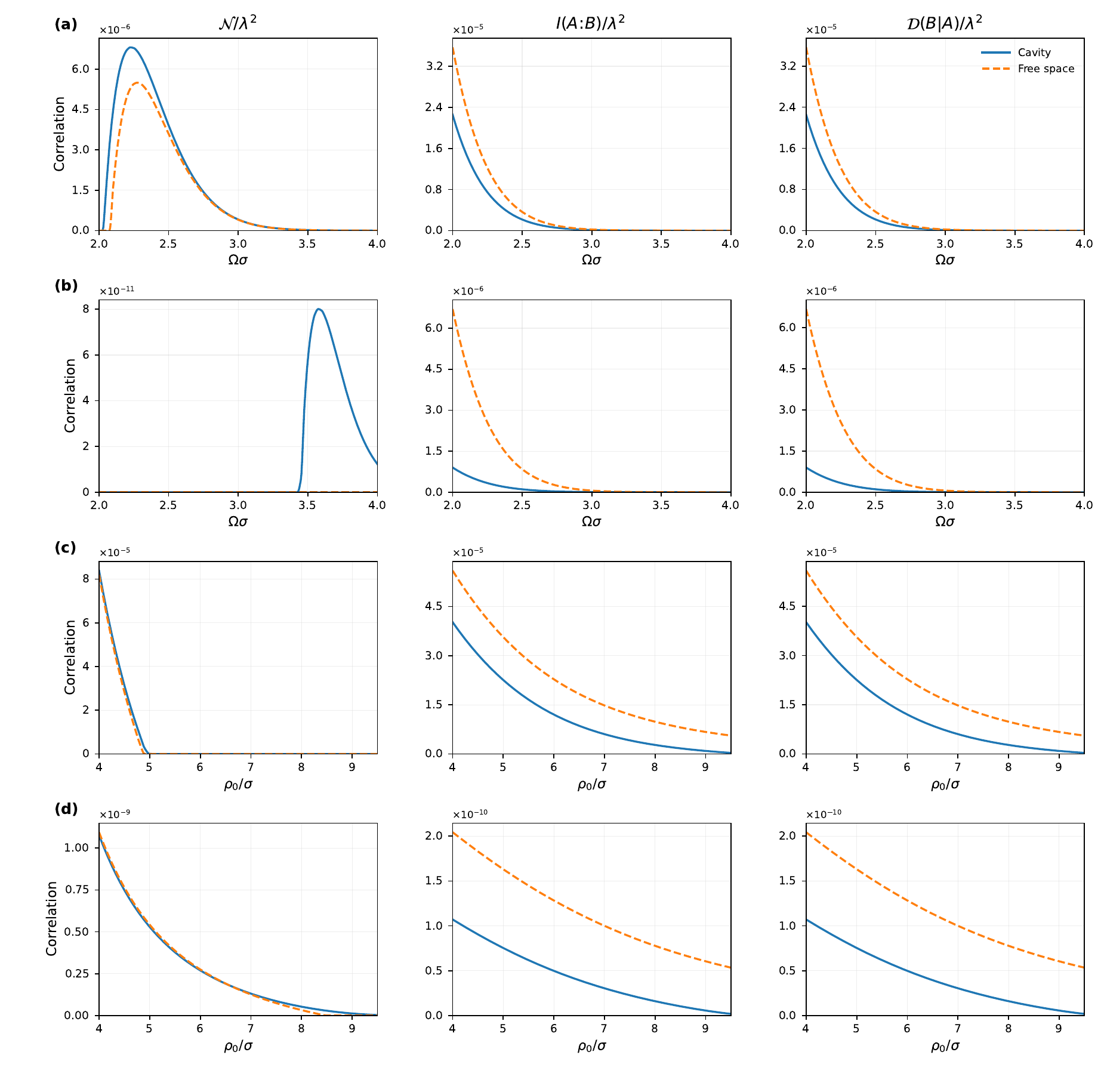}
    \caption{\it{\small
    Additional one-dimensional comparisons of the harvested
    correlations for the cylindrical cavity and free space.
    The three columns show the negativity
    $\mathcal{N}/\lambda^2$, mutual information
    $I(A\!:\!B)/\lambda^2$, and quantum discord
    $\mathcal{D}(B|A)/\lambda^2$, respectively.
    Panels (a)--(b) show the dependence on
    $\Omega\sigma$ at fixed
    $\rho_0/\sigma=5$ and $9$, respectively,
    for $\sigma/R=0.1$.
    Panels (c)--(d) show the dependence on
    $\rho_0/\sigma$ at fixed
    $\Omega\sigma=2$ and $4$, respectively,
    again for $\sigma/R=0.1$.
    In each panel, the solid curve corresponds to the
    cylindrical cavity and the dashed curve to free space.
    The mutual information and quantum discord are evaluated
    using the positive approximate state
    $\widetilde{\rho}_{AB}^{(2)}$ with the same numerical
    convention as in~\ref{fig:4-1dSlice}.
}    }
    \label{fig:app-extended-1d}
\end{figure*}

To further examine the parameter dependence of the
harvested correlations,~\ref{fig:app-extended-1d}
presents additional one-dimensional comparisons between
the cylindrical cavity and free space for different
detector separations and energy gaps. The results
complement~\ref{fig:4-1dSlice} and demonstrate
that the influence of cavity confinement on entanglement
harvesting depends sensitively on the detector
configuration.

For $\rho_0/\sigma=5$,~\ref{fig:app-extended-1d}(a)
shows that the cavity negativity exhibits a larger
maximum than its free-space counterpart, while both
quantities decrease rapidly at larger energy gaps.
At the larger separation $\rho_0/\sigma=9$,~\ref{fig:app-extended-1d}(b) reveals a
different behavior. The free-space negativity vanishes
throughout the displayed range, whereas the cavity
supports a narrow region of nonzero negativity
at sufficiently large $\Omega\sigma$.
Although this indicates an extension of the
entangled region due to confinement, the corresponding
negativity remains extremely small.

The separation-dependent results provide a
complementary perspective. At $\Omega\sigma=2$,~\ref{fig:app-extended-1d}(c) shows that
both cavity and free-space negativities vanish
beyond finite detector separations, with the
cavity exhibiting a slightly extended entangled
region. At $\Omega\sigma=4$,~\ref{fig:app-extended-1d}(d) shows that
positive cavity negativity persists to larger
separations than in free space, although its
magnitude decreases substantially.

In contrast to the parameter-dependent behavior
of negativity, the mutual information and quantum
discord remain smaller inside the cavity than
in free space throughout the displayed parameter
ranges. Both quantities decrease smoothly with
increasing energy gap or detector separation.
These entropic measures are evaluated using the
positive approximate reduced state
$\widetilde{\rho}_{AB}^{(2)}$, following the
numerical convention described in Section \ref{Section4}.

Taken together, these additional comparisons
demonstrate that cavity confinement can enhance
or extend the region of harvested entanglement
without a corresponding enhancement of mutual
information or quantum discord. The results
illustrate the distinct responses of the
correlation measures to boundary-induced
modifications of the field modes.

\bibliographystyle{cas-model2-names}

\begin{thebibliography}{99}

\bibitem{bell_1}
A.~Einstein, B.~Podolsky and N.~Rosen, {\it Can Quantum-Mechanical Description of Physical Reality Be Considered Complete}, Phys. Rev. \textbf{777} (1935) 

\bibitem{bell_2}
S.~Bell, {\it On the Einstein-Podolsky-Rosen paradox}, Physics\textbf{1}  195 (1964)

\bibitem{bell_3}
J.~F.~Clauser, M.~A.~Horne, A.~Shimony and R.~A.~Holt, {\it Proposed experiment to test local hidden-variable theories}, Phys.~Rev.~Lett.\textbf{23} 880 (1969) 

\bibitem{bell_4} 
  R.~F.~Werner,
  {\it Quantum states with Einstein-Podolsky-Rosen correlations admitting a hidden-variable model},
   Phys.\ Rev.\ A {\bf 40}, 4277 (1989)


\bibitem{WJHN}
W.~Tittel, J.~Brendel, H.~Zbinden and N.~Gisin,
{\it Violation of Bell inequalities by photons more than 10 km apart},
Phys. Rev. Lett. \textbf{81}, 3563-3566 (1998)
[arXiv:quant-ph/9806043 [quant-ph]].

\bibitem{DAC}
D. Salart, A. Baas, C. Branciard, N. Gisin and H. Zbinden, {\it Testing spooky action at a distance}, Nature 454, 861-864 (2008).

\bibitem{Reynaud:2001kc}
S.~Reynaud, A.~Lambrecht, C.~Genet and M.~T.~Jaekel,
{\it Quantum vacuum fluctuations},
Compt. Rend. Acad. Sci. Ser. IV Phys. Astrophys. \textbf{2}, no.9, 1287-1298 (2001) [arXiv:quant-ph/0105053 [quant-ph]].

\bibitem{Sakharov:1967pk}
A.~D.~Sakharov,
{\it Vacuum quantum fluctuations in curved space and the theory of gravitation},
Dokl. Akad. Nauk Ser. Fiz. \textbf{177}, 70-71 (1967).

\bibitem{Streeruwitz:1975wzf}
E.~Streeruwitz,
{\it Vacuum fluctuations of a scalar field in an Einstein universe}, Phys. Lett. B \textbf{55}, 93-96 (1975).

\bibitem{Menezes:2015veo}
G.~Menezes,
{\it Radiative processes of two entangled atoms outside a Schwarzschild black hole},
Phys. Rev. D \textbf{94}, no.10, 105008 (2016)
[arXiv:1512.03636 [gr-qc]].


\bibitem{Menezes:2015iva}
G.~Menezes and N.~F.~Svaiter,
{\it Radiative processes of uniformly accelerated entangled atoms},
Phys. Rev. A \textbf{93}, no.5, 052117 (2016)
[arXiv:1512.02886 [hep-th]].


\bibitem{Lindel:2023rfi}
F.~Lindel, A.~Herter, V.~Gebhart, J.~Faist and S.~Y.~Buhmann,
{\it Entanglement Harvesting from Electromagnetic Quantum Fields},
[arXiv:2311.04642 [quant-ph]].

\bibitem{Lima:2023pyt}
C.~Lima, E.~Patterson, E.~Tjoa and R.~B.~Mann,
{\it Unruh phenomena and thermalization for qudit detectors},
Phys. Rev. D \textbf{108}, no.10, 105020 (2023)
[arXiv:2309.04598 [quant-ph]].

\bibitem{Gallock-Yoshimura:2021xsy}
K.~Gallock-Yoshimura and R.~B.~Mann,
{\it Entangled detectors nonperturbatively harvest mutual information},
Phys. Rev. D \textbf{104}, no.12, 125017 (2021)
[arXiv:2109.07495 [quant-ph]].


\bibitem{Bueley:2022ple}
K.~Bueley, L.~Huang, K.~Gallock-Yoshimura and R.~B.~Mann,
{\it Harvesting mutual information from BTZ black hole spacetime},
Phys. Rev. D \textbf{106}, no.2, 025010 (2022)
[arXiv:2205.07891 [quant-ph]].

\bibitem{Quan:2026rmk}
M.~Quan, R.~Li and Z.~Zhao,
{\it Mutual information harvesting for circularly accelerated detectors},
Nucl. Phys. B \textbf{1026}, 117454 (2026)[arXiv:2604.12629 [quant-ph]].

\bibitem{Lin:2024roh}
F.~L.~Lin and S.~Mondal,
{\it Entanglement harvesting and quantum discord of alpha vacua in de Sitter space},
JHEP \textbf{08}, 159 (2024) [arXiv:2406.19125 [hep-th]].

\bibitem{Martin-Martinez:2015qwa}
E.~Martin-Martinez, A.~R.~H.~Smith and D.~R.~Terno,
Phys. Rev. D \textbf{93} (2016) no.4, 044001
doi:10.1103/PhysRevD.93.044001
[arXiv:1507.02688 [quant-ph]].

\bibitem{Pozas-Kerstjens:2015gta}
A.~Pozas-Kerstjens and E.~Martin-Martinez,
{\it Harvesting correlations from the quantum vacuum},
Phys. Rev. D \textbf{92} (2015) no.6, 064042
[arXiv:1506.03081 [quant-ph]].

\bibitem{Bhattacharya:2022ahn}
D.~Bhattacharya, K.~Gallock-Yoshimura, L.~J.~Henderson and R.~B.~Mann, {\it Extraction of entanglement from quantum fields with entangled particle detectors},
Phys. Rev. D \textbf{107}, no.10, 105008 (2023) [arXiv:2212.12803 [quant-ph]].

\bibitem{Suryaatmadja:2022quq}
C.~Suryaatmadja, R.~B.~Mann and W.~Cong,
{\it Entanglement harvesting of inertially moving Unruh-DeWitt detectors in Minkowski spacetime},
Phys. Rev. D \textbf{106}, no.7, 076002 (2022)
[arXiv:2205.14739 [quant-ph]].

\bibitem{Perche:2022ykt}
T.~R.~Perche, B.~Ragula and E.~Mart{\'\i}n-Mart{\'\i}nez,
{\it Harvesting entanglement from the gravitational vacuum},
Phys. Rev. D \textbf{108}, no.8, 085025 (2023)
[arXiv:2210.14921 [quant-ph]].

\bibitem{Barman:2022xht}
D.~Barman, S.~Barman and B.~R.~Majhi, {\it Entanglement harvesting between two inertial Unruh-DeWitt detectors from nonvacuum quantum fluctuations},
Phys. Rev. D \textbf{106}, no.4, 045005 (2022)
doi:10.1103/PhysRevD.106.045005
[arXiv:2205.08505 [gr-qc]].

\bibitem{deSLTorres:2023ujd}
B.~de S.~L.~Torres, K.~Wurtz, J.~Polo-G{\'o}mez and E.~Mart{\'\i}n-Mart{\'\i}nez,
{\it Entanglement structure of quantum fields through local probes},
JHEP \textbf{05} (2023), 058
[arXiv:2301.08775 [quant-ph]].

\bibitem{Bozanic:2023okm}
L.~Bozanic, M.~Naeem, K.~Gallock-Yoshimura and R.~B.~Mann, {\it Correlation harvesting between particle detectors in uniform motion},
Phys. Rev. D \textbf{108}, no.10, 105017 (2023)
[arXiv:2308.06329 [quant-ph]].

\bibitem{Perche:2023nde}
T.~R.~Perche, J.~Polo-G{\'o}mez, B.~de S.~L.~Torres and E.~Mart{\'\i}n-Mart{\'\i}nez,
{\it Fully relativistic entanglement harvesting},
Phys. Rev. D \textbf{109} (2024) no.4, 045018
[arXiv:2310.18432 [quant-ph]].

\bibitem{deRamon:2023qcp}
J.~de Ram{\'o}n, M.~Papageorgiou and E.~Mart{\'\i}n-Mart{\'\i}nez,
{\it Causality and signalling in noncompact detector-field interactions},
Phys. Rev. D \textbf{108} (2023) no.4, 045015
[arXiv:2305.07756 [quant-ph]].

\bibitem{Elghaayda:2023igv}
S.~Elghaayda and M.~Mansour,
{\it Entropy disorder and quantum correlations in two Unruh-deWitt detectors uniformly accelerating and interacting with a massless scalar field},
Phys. Scripta \textbf{98}, no.9, 095254 (2023)

\bibitem{Kaushal:2024zfi}
S.~Kaushal and S.~Bhattacharya, {\it Entanglement generation between Unruh{\textendash}DeWitt detectors in the de Sitter spacetime {\textemdash} Analysis with complex scalar fields},
Annals Phys. \textbf{482}, 170235 (2025)
[arXiv:2404.11931 [gr-qc]].

\bibitem{Ollivier:2001fdq}
H.~Ollivier and W.~H.~Zurek,
{\it Introducing Quantum Discord},
Phys. Rev. Lett. \textbf{88}, no. 1, 017901 (2001)
[arXiv:quant-ph/0105072 [quant-ph]].

\bibitem{Werlang:2009tvi}
T.~Werlang, S.~Souza, F.~F.~Fanchini and C.~J.~V.~Boas, {\it Robustness of quantum discord to sudden death},
Phys. Rev. A \textbf{80}, no.2, 024103 (2009)
[arXiv:0905.3376 [quant-ph]].


\bibitem{Teixido-Bonfill:2025wqb}
A.~Teixid{\'o}-Bonfill, X.~Dai, A.~Lupascu and E.~Mart{\'\i}n-Mart{\'\i}nez,
{\it Towards an experimental implementation of entanglement harvesting in superconducting circuits: Effect of detector gap variation on entanglement harvesting},
Phys. Rev. A \textbf{113}, no.4, 043732 (2026) [arXiv:2505.01516 [quant-ph]].

\bibitem{Lopez-Raven:2025ehf}
A.~Lopez-Raven, R.~B.~Mann and J.~Louko, {\it Quenched entanglement harvesting},
Phys. Rev. D \textbf{112}, no.8, 8 (2025) [arXiv:2506.07172 [gr-qc]].


\bibitem{Haroche2006}
S. Haroche and J.M. Raimond, {\it Exploring the Quantum: Atoms, Cavities, and Photon}, Oxford University Press (2006).


\bibitem{Walther2006}
H. Walther, B. T. H. Varcoe, B. G. Englert and T. Becker, {\it Cavity quantum electrodynamics}, Rep. Prog. Phys., \textbf{69}, 1325 (2006).

\bibitem{Raimond2001}
J. M. Raimond and M. Brune and S. Haroche, {\it Manipulating quantum entanglement with atoms and photons in a cavity}, Rev. Mod. Phys., \textbf{73}, 565 (2001).


\bibitem{Lochan:2019osm}
K.~Lochan, H.~Ulbricht, A.~Vinante and S.~K.~Goyal,
{\it Detecting Acceleration-Enhanced Vacuum Fluctuations with Atoms Inside a Cavity},
Phys. Rev. Lett. \textbf{125}, 241301 (2020)[arXiv:1909.09396 [gr-qc]].

\bibitem{Soda:2021sql}
B.~{\v{S}}oda, V.~Sudhir and A.~Kempf, {\it Acceleration-Induced Effects in Stimulated Light-Matter Interactions},
Phys. Rev. Lett. \textbf{128}, no.16, 163603 (2022)[arXiv:2103.15838 [quant-ph]].

\bibitem{Stargen:2021vtg}
D.~J.~Stargen and K.~Lochan,
Cavity Optimization for Unruh Effect at Small Accelerations,
Phys. Rev. Lett. \textbf{129} (2022) no.11, 111303
[arXiv:2107.00049 [gr-qc]].
 

\bibitem{Sahota:2026imu}
H.~S.~Sahota, S.~Kaushal and K.~Lochan,
{\it Cavity-controlled Inhibition of Decoherence in Accelerated Quantum Detectors}, [arXiv:2604.02422 [gr-qc]].

\bibitem{Liu:2020jaj}
Z.~Liu, J.~Zhang and H.~Yu, {\it Entanglement harvesting in the presence of a reflecting boundary},
JHEP \textbf{08}, 020 (2021) [arXiv:2101.00114 [quant-ph]].

\bibitem{Li:2024dvs}
R.~Li and Z.~Zhao, {\it Entanglement harvesting of circularly accelerated detectors with a reflecting boundary},
JHEP \textbf{03}, 185 (2025) [arXiv:2401.16018 [quant-ph]].

\bibitem{Barman:2023wkr}
D.~Barman and B.~R.~Majhi, {\it Are multiple reflecting boundaries capable of enhancing entanglement harvesting?},
Phys. Rev. D \textbf{108}, no.8, 8 (2023)[arXiv:2306.09943 [gr-qc]].

\bibitem{Ma:2024tyn}
C.~Ma and Z.~Zhao, {\it Effect of environment-induced interatomic interaction on entanglement generation for uniformly accelerated atoms with a boundary},
Phys. Lett. B \textbf{867}, 139593 (2025) [arXiv:2410.21056 [quant-ph]].

\bibitem{Salomaa:2026xzn}
S.~Salomaa, E.~Keski-Vakkuri and S.~Nadal-Gisbert,
{\it Bipartite entanglement harvesting with multiple detectors},
[arXiv:2604.13869 [quant-ph]].

\bibitem{Wu:2026woq}
S.~M.~Wu, X.~Y.~Jiang, X.~Y.~Yu, Z.~Liu and X.~L.~Huang, {\it Reflecting boundary induced modulation of tripartite coherence harvesting},
JHEP \textbf{05}, 123 (2026) [arXiv:2601.21240 [quant-ph]].


\bibitem{Huang:2025gme}
X.~L.~Huang, X.~Y.~Jiang, Y.~X.~Wang, S.~Y.~Liu, Z.~Wang and S.~M.~Wu,
{\it Can boundary configuration be tuned to optimize directional quantum steering harvesting?},
JHEP \textbf{09}, 023 (2025) [arXiv:2506.18734 [quant-ph]].


\bibitem{Mendez-Avalos:2022obb}
D.~Mendez-Avalos, L.~J.~Henderson, K.~Gallock-Yoshimura and R.~B.~Mann,
{\it Entanglement harvesting of three Unruh-DeWitt detectors},
Gen. Rel. Grav. \textbf{54}, no.8, 87 (2022)
[arXiv:2206.11902 [quant-ph]].

\bibitem{Mukherjee:2023bnt}
A.~Mukherjee, S.~Gangopadhyay and A.~S.~Majumdar,
{\it Fulling-Davies-Unruh effect for accelerated two-level single and entangled atomic systems},
Phys. Rev. D \textbf{108}, no.8, 085018 (2023)
[arXiv:2305.08867 [quant-ph]].

\bibitem{Strohle:2026gdr}
J.~Str{\"o}hle and N.~Momcilovic,
{\it Entanglement harvesting in the presence of cavities},
[arXiv:2601.16698 [quant-ph]].



\bibitem{NielsenChuang} 
	M.~A.~Nielsen and I.~L.~Chuang (2010), {\it Quantum Computation And Information Theory} (Cambridge university press, UK)

\bibitem{Vidal:2002zz} 
  G.~Vidal and R.~F.~Werner,
  {\it Computable measure of entanglement},
  Phys.\ Rev.\ A{\bf 65}, 032314 (2002)
  [arXiv:quant-ph/0102117].

\bibitem{Plenio:2005} 
  M.~B.~Plenio,
  {\it Logarithmic negativity: a full entanglement monotone that is not convex},
  Phys.\ Rev.\ Lett.{\bf 95}, 090503 (2005).
 
  \bibitem{Horodecki:1997vt}
P.~Horodecki,
{\it Separability criterion and inseparable mixed states with positive partial transposition},
Phys. Lett. A\textbf{232}, 333 (1997)
[arXiv:quant-ph/9703004 [quant-ph]].

\bibitem{Datta:2009wug}
A.~Datta,
{\it Quantum discord between relatively accelerated observers},
Phys. Rev. A \textbf{80}, no.5, 052304 (2009)
[arXiv:0905.3301 [quant-ph]].

\bibitem{Martin-Martinez:2012ysv}
E.~Martin-Martinez, M.~Montero and M.~del Rey,
{\it Wavepacket detection with the Unruh-DeWitt model},
Phys. Rev. D \textbf{87}, no.6, 064038 (2013)
[arXiv:1207.3248 [quant-ph]].

\bibitem{gradshteyn2007}Gradshteyn, I., Ryzhik, I., Jeffrey, A. \& Zwillinger, D. Table of Integrals, Series, and Products. (Elsevier/Academic Press,2007)

\end{thebibliography}
\bibdata{outputNotes,outputNotes : ,cas-refs}

\end{document}